\documentclass[preprint]{elsarticle}
\journal{Acta Astronautica}
\myfooter[L]{Accepted for publication in Acta Astronautica}
\date{}

\usepackage{lineno,hyperref,eurosym}
\modulolinenumbers[5]
\usepackage{color,xcolor}
\usepackage{moresize}
\usepackage{natbib}

\usepackage{amsmath, mathtools, amssymb, bm}
\usepackage{algpseudocode}
\usepackage{algorithm}
\usepackage{siunitx} 

\usepackage{}

\usepackage{comment}

\usepackage{graphicx}
\usepackage{booktabs}
\usepackage{tabu}
\graphicspath{{images/}}
\usepackage{caption}
\usepackage{subcaption}
\usepackage{url}
\usepackage{setspace}
\usepackage{fancyhdr}
\usepackage{hyperref}
\usepackage{diagbox}
 \usepackage{xcolor}
\usepackage{tabularx}
\usepackage{booktabs}
\usepackage{threeparttable}
\usepackage{url}
\usepackage{multirow} 
\usepackage{booktabs}
\usepackage{makecell}
\usepackage{natbib}
\let\cite\citep

\begin{document}

\begin{frontmatter}

\title{The EMA Justitia Probe:\\ Terminator Frame-Based Deployment Orbit Selection and Surface Coverage Analysis}
\author[label1,label2]{S. Biktimirov*}
\author[label1]{F. Alnaqbi}
\author[label1]{A. Ivanov}

\address[label1]{Propulsion and Space Research Center, Technology Innovation Institute, Abu Dhabi (United Arab Emirates)}
\address[label2]{Department of Aerospace Engineering, Khalifa University of Science and Technology, Abu Dhabi (United Arab Emirates)}

\cortext[]{Corresponding author: biktimirovshamil@gmail.com (S.~Biktimirov)}

\begin{abstract}

The Emirates Mission to the Asteroid Belt probe is designed to perform asteroid imaging during an uncontrolled ballistic descent to (269) Justitia. The mission design is largely shaped by the probe technical requirements and by the operational constraints associated with the main spacecraft proximity operations. In particular, the communication architecture does not allow updated asteroid parameters or final deployment orbit information to be uploaded to the probe during the proximity operations phase. At the same time, the range of possible deployment orbits is constrained by the operational and safety requirements of the main spacecraft -- the deployment orbit shall remain close to the terminator plane to avoid eclipse conditions, while its periapsis altitude shall remain sufficiently high to maintain safe main spacecraft operations. In addition, uncertainty in the asteroid gravity field prevents the final deployment orbit from being selected before the proximity operations phase. These limitations require a mission design that provides suitable illumination, attitude determination and control, and payload pointing conditions throughout descent without prior knowledge of the exact deployment trajectory. The EMA probe mission concept of operations is delineated, with particular emphasis on an attitude control approach. The approach utilizes the near-terminator deployment orbit geometry and continuous Sun visibility. It includes a single-axis Sun-spin mode providing periodic observation opportunities and a three-axis asteroid-pointing mode providing continuous asteroid visibility. Based on this concept, formal requirements are derived for deployment orbits that provide asteroid observation opportunities throughout the ballistic descent. These requirements form the basis for the final descent trajectory selection and represent a mission-level outcome of the EMA probe design. A formal deployment orbit selection method is then developed using the terminator reference frame and a minimum boresight deviation margin. The method identifies families of orbit orientations relative to the terminator plane that satisfy illumination and payload observation requirements. Reference orbits are selected and verified through mission geometry analysis and trajectory propagation using a polyhedron gravity model. Finally, asteroid surface coverage is evaluated for uncontrolled tumbling, Sun-spin, and three-axis asteroid-pointing modes. The results characterize the expected imaging performance and demonstrate the feasibility of the EMA probe concept for close-proximity imaging of a relatively large asteroid. The proposed approach can also support future ballistic probe missions subject to illumination, observability, main-spacecraft orbit constraints, and limited onboard knowledge of the final descent trajectory.

\end{abstract}


\end{frontmatter}

\section{Introduction} 
\label{sec:introduction}

The work presents the Emirates Mission to the Asteroid Belt (EMA) probe mission concept and focuses on deployment-orbit selection based on the probe requirements, the current main spacecraft (S/C) proximity operations strategy, and the system design described in~\cite{ema_SE}. The general EMA mission design, including the interplanetary trajectory, flybys of six asteroids, rendezvous with asteroid (269) Justitia, and the main spacecraft proximity operations, is detailed in~\cite{almazmi2023scienceB,alhameli2023mission,parkerEmaMissionDesign,parkerproximity}. The EMA probe is a deployable spacecraft element of the main S/C, developed in the United Arab Emirates with Technology Innovation Institute (TII) leading its development. The probe is planned to be deployed from the main spacecraft orbit around Justitia and perform scientific observations during its descent to the asteroid surface.
According to the current requirements, the probe shall not include a propulsion system or deployable structures. Therefore, after separation from the main S/C, the probe performs an uncontrolled ballistic descent, while its active attitude determination and control system provides the required payload camera pointing. Thus, the probe ConOps and scientific observations depend on the deployment orbit and deployment conditions.
The present study considers the probe descent phase, attitude control approach, deployment orbit selection, and camera observation performance. Reference descent trajectories and asteroid surface coverage are analyzed to evaluate the proposed mission design.
To understand the approaches used for the design of small-body missions with deployable spacecraft, previous and currently planned missions involving landers, CubeSats, and close-proximity probes are reviewed below.

Several successful small-body missions have involved deployable spacecraft~\cite{ULAMEC2014460}, and this direction continues to be actively explored through new planned missions. The following missions are considered to investigate the aspects of small body missions with deployable spacecraft taking into account system-level design, descent dynamics, attitude determination and control, among other aspects. Together, these missions demonstrate a range of approaches for small-body exploration, from thruster-less landers with uncontrolled rotational motion performing in-situ surface investigation, to thruster-assisted sample collection with active attitude control, and to a nanosatellite performing remote observations, all relying on precise orbital and attitude dynamics modeling in small body environments.

The 100 kg Philae lander, developed by an international consortium led by DLR and carried by the ESA's Rosetta spacecraft, explored Jupiter-family comet 67P/Churyumov-Gerasimenko, a bilobate shape body with a large lobe size of $4.1 \times 3.3  \times 1.8$~km and small lobe size of $2.6 \times 2.3 \times 1.8$~km, and 12.4 h rotation period. On November 12th, 2014, the lander separated from Rosetta at 22.5 km from the center of the comet with expected descent time of approximately 7 hours and touchdown velocity of 0.95 m/s. Prior to deployment, a detailed investigation of the comet with reconstruction of its gravitational parameter $\mu = 667~ m^3/s^2$, gravity field, and surface topography was performed. The descent duration was close to the predicted value, with a difference of 50~s and the first touchdown point was approximately 120~m from the nominal target site. However, due to a double failure of the harpoons and thruster the lander bounced from the surface and after around 1~h 50~min of bouncing came to rest approximately 1~km away from the target location in a poorly illuminated area. The lander performed valuable scientific observations for 5 comet days (57 h) until its batteries were completely discharged and it stopped communicating with Rosetta. The aspects of the mission related to flight dynamics, landing site selection, and scientific instruments of the spacecraft are presented in~\cite{Jurado2016,Witte2016, biele2008capabilities, Biele2015}.

The Hayabusa-2 mission MASCOT lander~\cite{ho2017mascot} performed a 17-hour-long in-situ exploration of near-Earth asteroid Ryugu with four scientific instruments: a hyperspectral microscope MicrOmega, radiometer MARA~\cite{grott2017mascot}, magnetometer MASMag~\cite{hervcik2017mascot}, and camera MASCam~\cite{jaumann2017camera}. It benefited from the heritage of the Philae scientific payload. The lander was released in the weak-gravity environment of the approximately 900-m diameter asteroid with a spring deployment mechanism at a low altitude of ~50 m and performed its first touchdown in around 6 minutes, while it took approximately half an hour to reach its first resting position where the observation started. The lander was equipped with a swing arm to change its orientation on the surface and perform movement on the surface to explore different sites. A detailed modeling of MASCOT landing taking into account surface interaction is presented in~\cite{Thuillet2018}. The post-mission interpretation of the landing results is demonstrated in~\cite{ho2021mascot, Thuillet2021}. Images acquired on the surface of asteroid Ryugu and their scientific interpretation are presented in~\cite{jaumannMascot}.

LICIAcube, a 6U nanosatellite presents another type of deployable spacecraft that performs an asteroid flyby. LICIAcube was deployed from DART spacecraft to perform a flyby of the Didymos--Dimorphos binary asteroid system and capture high-resolution images of the ejecta plume after the kinetic impact of the DART spacecraft~\cite{DOTTO2021105185}. LICIAcube’s scientific payload included two optical cameras: LEIA (a narrow-field high-resolution panchromatic camera) and LUKE (a wide-field RGB camera for color analysis) designed to characterize ejecta morphology, dynamics, and particle distribution \cite{DOTTO2021105185}. LICIACube was deployed 15 days before the impact. Equipped with an onboard propulsion system, it then performed a maneuver to follow behind the DART spacecraft and observe its impact with the asteroid Dimorphos. Observations were acquired from 71 seconds before the impact until 320 seconds afterwards~\cite{10521222, dotto2024dimorphos}.

Several ongoing and planned missions further demonstrate the variety of spacecraft architectures used for asteroid and small-body exploration, including sample-return spacecraft, landers, and deployable CubeSats. China's Tianwen-2~\cite{tianwen2_2025}, launched in May 2025, aims to return samples from the near-Earth asteroid 2016 HO3 also known to be a quasi-satellite of the Earth and to study the main-belt comet 311P/PANSTARRS. Various aspects of Tianwen-2 mission design including proximity dynamics and control modeling, system-level modeling for asteroid sampling and asteroid 3D modeling simulation are presented in~\cite{tianwen2_dynamics_control, Zhang2025, tianwen2_3dmodelling}. Japan's Next Generation Small-body Sample Return (NGSR) mission is intended to return samples from a comet nucleus using a Deep Space Orbital Transfer Vehicle (DSOTV), which supports proximity operations such as in-situ analysis and bistatic radar measurements. Critical technologies for the lander and its instruments are currently under development \cite{Saiki2025}. The European RAMSES mission~\cite{MORELLI20244241} is proposed to study near-Earth asteroid 99942 Apophis and contribute to planetary defense research. Recently, ESA announced the CubeSat lander Don Quijote to operate and perform measurements on the surface of the asteroid~\cite{ESA_DonQuijote_2026}. The ESA Hera mission is the European contribution to the AIDA planetary defense campaign~\cite{michel2018european} and is aimed at the detailed characterization of the Didymos–Dimorphos binary asteroid system after the DART kinetic impact experiment. A distinctive feature of the mission is the deployment of the two CubeSats, Milani and Juventas, which will perform autonomous close-proximity operations around the asteroid system~\cite{ferrari2021preliminary, michel2022esa}. Rather than landing on the surface or executing a single flyby, the CubeSats are designed to operate on bounded low-altitude trajectories and proximity orbits in the extremely weak and irregular gravity environment of Dimorphos and Didymos. Both CubeSats are equipped with propulsion and ADCS subsystems to support autonomous operations and precise attitude pointing during the science phase. The CubeSats will carry complementary payloads for the investigation of the asteroid surface, dust environment, gravity field, and internal structure using imaging, spectroscopic, and low-frequency radar instruments.

Previous deployable landers mainly targeted relatively small bodies with characteristic dimensions of about one kilometer. In such weak-gravity environments, ballistic descent can result in a relatively low touchdown velocity, while the following surface motion is often similar to bouncing or hopping. Small velocity changes can also significantly modify the spacecraft trajectory near such bodies. In contrast, (269) Justitia has a diameter of approximately 60~km and a more significant and irregular gravity field. Therefore, the EMA probe moves along a ballistic descent orbit and reaches the asteroid surface with a velocity of several tens of meters per second. The main scientific observations shall consequently be performed during descent rather than after touchdown.

The present work has been partially presented in~\cite{ema_biktimirov} and continues the analysis performed in~\cite{alnaqbi_MPC_ema_lander,Biktimirov2024}. The EMA probe design and systems engineering aspects are presented in~\cite{ema_SE}. The previous studies considered lander concepts that could control the descent orbit using a thruster and model predictive control, or decrease the touchdown velocity by separating a detachable part with a spring mechanism. However, according to the updated mission requirements, the probe shall not include a propulsion system or any deployable structures. Therefore, uncontrolled ballistic descent remains the only possible option. At the same time, the probe is planned to be equipped with an active attitude control system to improve asteroid imaging opportunities during descent. Since the spacecraft is neither intended to perform a soft landing and conduct in-situ observations nor currently considered as an impactor supporting the scientific measurements of the main spacecraft, the concept was renamed from the EMA lander to the EMA probe. Compared with the preliminary study presented in~\cite{ema_biktimirov}, the present work introduces the terminator frame-based method for deployment orbit attitude selection based on illumination and camera-pointing constraints. It also extends the analysis by evaluating asteroid surface coverage for uncontrolled tumbling, Sun-spin, and asteroid-pointing attitude control modes. The Lyapunov function-based attitude control laws utilized for the development of the probe attitude control algorithms are not introduced as a new contribution and therefore their description is moved to the~\ref{app: attitude control algorithms} section. They are used in the work to define the mission ConOps, to derive the corresponding deployment orbit requirements, and to perform the asteroid surface coverage analysis for the controlled attitude motion dynamics.

These mission properties make deployment orbit selection an important mission design problem. The deployment orbit shall be compatible with the main spacecraft proximity operations and, at the same time, satisfy the requirements related to descent duration, asteroid illumination, attitude control, and payload camera observations. Previous studies do not provide an approach that considers these requirements together for a ballistic probe deployed near a relatively large and irregular asteroid. Thus, deployment orbit selection, attitude control approach, and imaging performance cannot be considered independently.

The present study derives formal requirements for the deployment orbit based on the probe ConOps and the current representation of the main spacecraft proximity orbits using Keplerian elements~\cite{parkerproximity}. A method is developed to select the deployment orbit orientation that is suitable for the proposed attitude control approach and provides asteroid observation conditions during the descent. The selected deployment orbits are further evaluated using reference trajectory simulations and asteroid surface coverage analysis.

The paper is structured as follows. Section~\ref{sec:reference frames and notations} introduces reference frames, main notations, and equations of orbital and angular motion dynamics used in the study. Section~\ref{sec:Justitia properties} summarizes currently available data on the target asteroid 269 Justitia. Section~\ref{sec: Lander description} presents the EMA probe mission concept, probe technical specifications and outlines the attitude control approach. Section~\ref{sec:deployment orbit selection} delineates probe deployment orbit requirements, proposes the terminator frame method that is used for the orbit selection and analysis, and finally performs Keplerian orbit dynamics analysis and selects candidate orbits for further detailed investigation of probe dynamics and control. Asteroid surface coverage performance is analyzed in Section~\ref{sec: Surface Coverage Analysis}. The outcomes of the study are summarized in Conclusion.

\section{Reference Frames, Notations, and Equations of Motion}
\label{sec:reference frames and notations}

\subsection{Reference Frames and Notations}
A vector $\mathbf{r} \in \mathbb{R}^3$ is treated as a geometric object independent of its coordinate representation. An orthonormal reference frame $\mathcal{F}^{\mathcal{A}}$ is defined by an origin $O$ and an ordered orthonormal basis $\mathcal{F}^{\mathcal{A}} = \{\mathbf{e}_1^{\mathcal{A}}, \mathbf{e}_2^{\mathcal{A}}, \mathbf{e}_3^{\mathcal{A}} \}$, where $\mathbf{e}_i^{\mathcal{A}}$ are mutually orthonormal basis vectors fixed in the frame. The coordinate representation of a vector $\mathbf{r}$ in frame $\mathcal{F}^{\mathcal{A}}$ is denoted by
$\mathbf{r}^{\mathcal{A}} =
[r_1^{\mathcal{A}}, r_2^{\mathcal{A}}, r_3^{\mathcal{A}}]^{\top}$, such that  $ \mathbf{r} = \sum_{i=1}^{3} r_i^{\mathcal{A}} \mathbf{e}_i^{\mathcal{A}}$.

Superscripts attached to vectors denote coordinate representation in the corresponding frame, while superscripts attached to basis vectors indicate the frame to which the basis belongs. The hat notation denotes a normalized vector $\hat{\mathbf{r}} = \frac{\mathbf{r}}{\|\mathbf{r}\|}$.

\begin{figure}[h!]
 \begin{center}
  \includegraphics[width=0.7\textwidth]{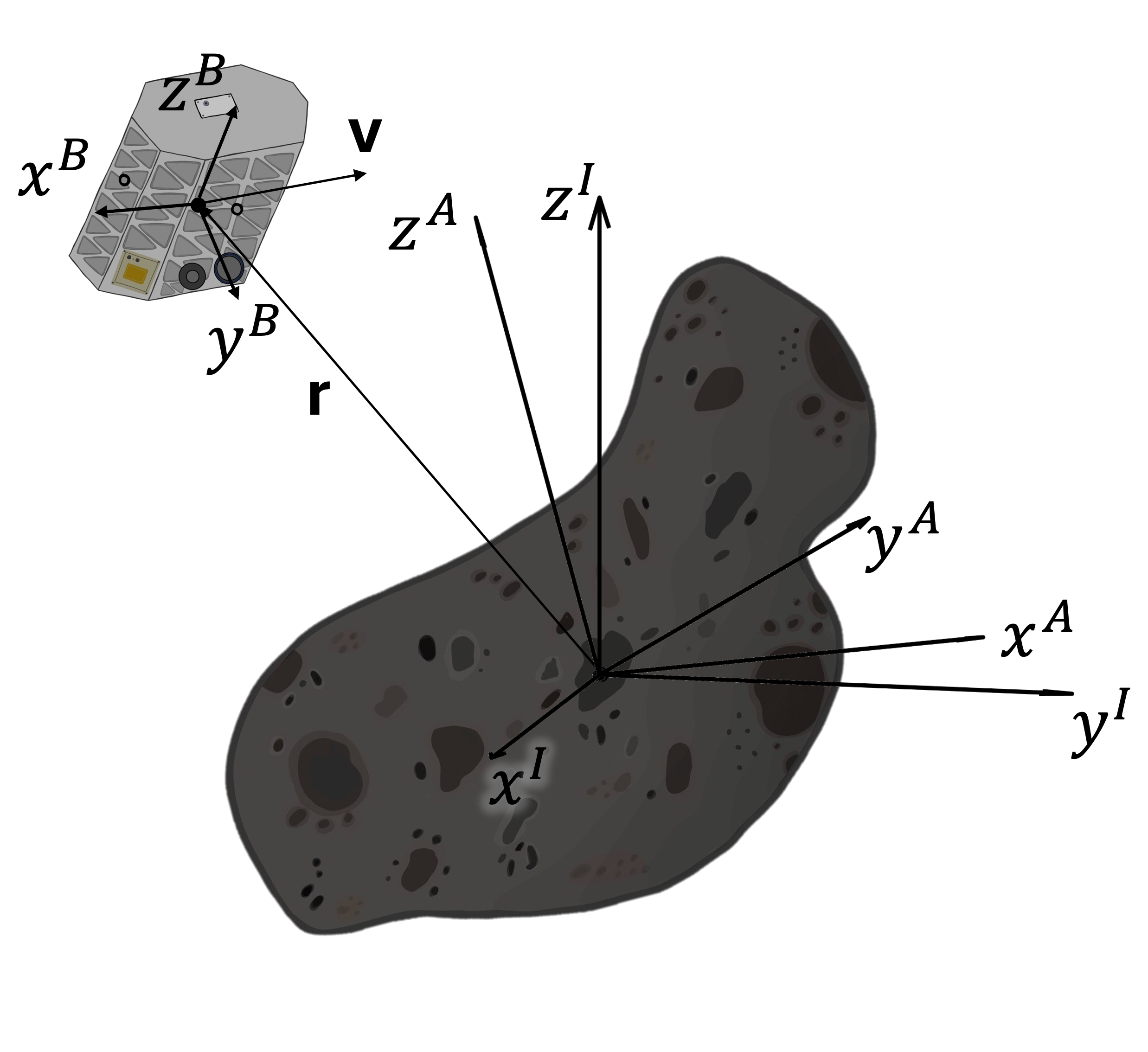}
  \caption{Reference frames}
  \label{fig:reference_frames}
   \end{center}
\end{figure}

The following reference frames to describe the motion of the probe are used. Figure~\ref{fig:reference_frames} illustrates the reference frames. Let us note that in figures, the axes of frame $\mathcal{F^A}$ are labeled as $x^\mathcal{A}$, $y^\mathcal{A}$, $z^\mathcal{A}$, corresponding to the basis vectors $\mathbf{e}_1^{\mathcal{A}}$, $\mathbf{e}_2^{\mathcal{A}}$, $\mathbf{e}_3^{\mathcal{A}}$, respectively.

\begin{enumerate}

    \item The orthonormal reference frame $\mathcal{F^I}$ is the asteroid-centered inertial ecliptic frame. denoted by $\mathcal{F^I} = \{ \mathbf{e}_1^{\mathcal{I}}, \mathbf{e}_2^{\mathcal{I}}, \mathbf{e}_3^{\mathcal{I}} \}$. The frame has its origin at the asteroid's center of mass, its $\mathbf{e}_1^{\mathcal{I}}$ points toward the Sun at vernal equinox, $\mathbf{e}_3^{\mathcal{I}}$ is aligned with normal to the ecliptic plane and pointed to the Earth's orbit angular momentum vector direction, $\mathbf{e}_2^{\mathcal{I}}$ axis completes the reference frame into a right-handed triad
    
    \item The orthonormal reference frame $\mathcal{F^T} = \{ \mathbf{e}_1^{\mathcal{T}}, \mathbf{e}_2^{\mathcal{T}}, \mathbf{e}_3^{\mathcal{T}} \}$ is the terminator frame used to design and assess the probe deployment orbit. The frame has origin at the asteroid center of mass and is defined as follows
\begin{equation}\label{eq:terminator frame}
    \mathbf{e}_3^{\mathcal{T}} = \hat{\mathbf{r}}_{\mathrm{Sun}}, \quad 
\mathbf{e}_1^{\mathcal{T}} = \frac{\mathbf{e}_3^{\mathcal{T}}\times\mathbf{e}_3^{\mathcal{I}}}{\left\| \mathbf{e}_3^{\mathcal{T}} \times \mathbf{e}_3^{\mathcal{I}} \right\|},
\quad \mathbf{e}_2^{\mathcal{T}} = \mathbf{e}_3^{\mathcal{T}} \times \mathbf{e}_1^{\mathcal{T}}
\end{equation}

\noindent let us note that z-axes of the terminator frame $\mathcal{F^T}$ and inertial frame $\mathcal{F^I}$ are used here to define the x-axis of the terminator frame, also $\left\| \mathbf{e}_3^{\mathcal{T}}\times\mathbf{e}_3^{\mathcal{I}} \right\| > 0$ for the considered mission geometry

    \item The orthonormal reference frame $\mathcal{F^A} = \{ \mathbf{e}_1^{\mathcal{A}}, \mathbf{e}_2^{\mathcal{A}}, \mathbf{e}_3^{\mathcal{A}} \}$ is the non-inertial asteroid-fixed reference frame
    
    \item The orthonormal reference frame $\mathcal{F^B} = \{ \mathbf{e}_1^{\mathcal{B}}, \mathbf{e}_2^{\mathcal{B}}, \mathbf{e}_3^{\mathcal{B}} \}$ is the non-inertial probe body-fixed frame
    
\end{enumerate}

The direction cosine matrix (DCM) $\mathbf{C}^{IJ}$ or a unit quaternion $\mathbf{q}^{IJ}$ transforming coordinates from frame $J$ to frame $I$ follows notation 

\begin{subequations}
\begin{align}
\label{eq:coordinate transformation definition}
\mathbf{r}^I &= \mathbf{C}^{IJ}\mathbf{r}^J \\
 \mathbf{r}^I &= \mathbf{q}^{IJ} \circ \mathbf{r}^J  \circ \tilde{\mathbf{q}}^{IJ},
\end{align}    
\end{subequations}

\noindent where "$\circ$" denotes quaternion multiplication, and $\tilde{\mathbf{q}}$ is quaternion conjugate. The position vector in the equation is considered as a quaternion with zero scalar part.

The transformation between $\mathcal{F^I}$ and $\mathcal{F^A}$ frames is defined as follows~\cite{damit}:

\begin{subequations}
\label{eq:coordinate transformation}
\begin{align}
\mathbf{r}^{\mathcal{I}} &= \mathbf{C}^{\mathcal{IA}}(t) \mathbf{r}^{\mathcal{A}}, \\
\mathbf{C}^{\mathcal{IA}}(t) &= \mathbf{R}_z(\lambda) \mathbf{R}_y(\frac{\pi}{2} - \theta) \mathbf{R}_z(\phi_0 + \frac{2\pi}{P}(t - t_0)) \label{eq:asteroid angles}
\end{align}
\end{subequations}

\noindent where $P$ is the asteroid rotation period, angles $\lambda, \theta$ define asteroid spin axis, angle $\phi_0$ defines asteroid rotation angle at time $t_0$, $\mathbf{R}_z$ and $\mathbf{R}_y$ are SO(3) matrices describing rotations around z- and y-axes, respectively.

\subsection{Equations of Motion}

The following equation describing orbital motion dynamics in the frame $\mathcal{F^I}$ is used:

\begin{equation} \label{eq:equation of motion}
    \ddot{\mathbf{r}} = - \nabla U (\mathbf{r}),
\end{equation}

\noindent where  $U$ is the asteroid gravity field potential. Solar radiation pressure is neglected due to the short descent duration and relatively weak solar perturbation at the operating distance ($\sim 3 AU)$. While a point-mass gravity field $U=-\frac{\mu}{r}$ is used for the Keplerian orbit selection, a uniform density polyhedron gravity model~\cite{werner1996exterior} is adopted to assess the selected reference trajectories. The point-mass approximation provides a clear geometrical interpretation of the orbit size, descent time, phase angle, and camera-visibility constraints, whereas the polyhedron model takes into account the irregular shape of Justitia which is important to consider in the close-proximity region, especially. 
The polyhedron gravity model provides analytical expressions for the gravitational potential and acceleration, and is also used to determine whether the probe position is outside or inside the asteroid shape for touchdown condition detection.  The corresponding equations for the potential, acceleration, and Laplacian-based inside/outside polyhedron condition are summarized in~\ref{app: polyhedron gravity model}.

Let us note that the polyhedron model considers satellite position vector represented in asteroid body-fixed frame $\mathcal{F^A}$, while the gravity attraction vector $\nabla U(\mathbf{r}^{\mathcal{A}})$ should be converted into the inertial frame $\mathcal{F^I}$ using the DCM matrix $\mathbf{C}^{\mathcal{IA}}(t)$ (see Eq.~\ref{eq:coordinate transformation}) for the orbital motion dynamics equation (see Eq.~\ref{eq:equation of motion}).

The equations describing angular motion kinematics and dynamics are Poisson's and Euler's equations, respectively:
\begin{subequations}
    \begin{align}\label{eq:equations of motion}
    \dot{\mathbf{q}}^{\mathcal{IB}} &= \frac{1}{2} \mathbf{q}^{\mathcal{IB}} \circ \boldsymbol{\omega}^{\mathcal{B}}, \\
    \dot{\boldsymbol{\omega}}^\mathcal{B} &= \mathbf{J}^{-1} (\mathbf{M}^\mathcal{B} - \boldsymbol{\omega}^\mathcal{B} \times \mathbf{J}\boldsymbol{\omega}^\mathcal{B}),
\end{align}
\end{subequations}

\noindent where $\mathbf{q}^{\mathcal{IB}}$ is a unit quaternion representing body-fixed axes attitude with respect to the inertial frame, $\boldsymbol{\omega}^\mathcal{B}$ is the probe's angular velocity projected to the body-fixed axes, $\mathbf{J}$ is the inertia tensor given in probe's body-fixed axes, $\mathbf{M}^\mathcal{B}$ denotes torque projected to the body-fixed axes including control torque and perturbations. External disturbance torques (e.g., gravity-gradient and solar radiation pressure) are neglected due to the short descent duration.

\section{Physical Properties of Asteroid 269 Justitia}
\label{sec:Justitia properties}

Asteroid (269) Justitia is among the reddest known objects in the inner Solar System, classified as a D-type or newly defined Z-class main-belt asteroid with spectra typical of trans-Neptunian objects ~\cite{marciniak2025thermophysical, buie_justitia_2025}. Following stellar occultation events in 2023 Justitia’s size, shape, and spin axis estimates were improved – equivalent sphere diameter $58\pm2$~km, shape model in a form of polyhedron is available in DAMIT catalogue~\cite{damit}. 

Justitia density $\rho$ is currently assumed to be in the range 1--2 $\mathrm{g/cm^3}$ as considered in the analysis of the proximity operations of the main spacecraft - MBR Explorer~\cite{parkerproximity}. The bulk density of D-type asteroids typically is within a range 1.0–1.6 $\mathrm{g/cm^3}$, consistent with primitive, porous, volatile-rich compositions~\cite{Carry2012, grav2012wise}. Considering the spectral similarity of Justitia to trans-Neptunian objects and its likely high macroporosity, a conservative range $\rho$ = 1–2 $\mathrm{g/cm^3}$ is adopted to bound gravitational parameter uncertainty. The high-density case represents a conservative scenario to bound worst-case gravitational acceleration.

Following prior studies of MBR Explorer proximity operations~\cite{parkerproximity}, three Justitia size and mass cases are considered in the analysis with pairs ($R_{\mathrm{eq}}$, $\rho$): (26.6; 1), (28; 1.3), (29.4; 2) km, $\mathrm{g/cm^3}$. The gravitational parameter $\mu$ for asteroid Justitia with equivalent radius $R_{\mathrm{eq}}$ and density $\rho$ is as follows:
\begin{equation}\label{eq:grav parameter}
   \mu = GM = G \rho V = G  \rho  \frac{4}{3} \pi  R_{\mathrm{eq}}^3, 
\end{equation}

\noindent yielding $\mu = 5.282 \times 10^{6}; 7.98 \times 10^{6}; 14.16 \times 10^{6} ~ \mathrm{m^3/s^2}$. 
Table~\ref{tab: Justitia parameters case} summarizes the Justitia size, density, and gravitational parameter cases that will be used for further analysis in the study.

\begin{table}[h]
\centering
\caption{Justitia cases}
\label{tab: Justitia parameters case}

\begin{tabular}{lccc}
\toprule
Case & $R_{\mathrm{eq}}$ [km] & $\rho$ [g/cm$^3$] & $\mu$ [m$^3$/s$^2$] \\
\midrule

Low $\mu$  & 26.6 & 1   & $5.282 \times 10^6$ \\
Mid $\mu$  & 28   & 1.3 & $7.98 \times 10^6$ \\
High $\mu$ & 29.4 & 2   & $14.16 \times 10^6$ \\

\bottomrule
\end{tabular}

\end{table}

\begin{figure}[h!]
 \begin{center}
  \includegraphics[width=1\textwidth]{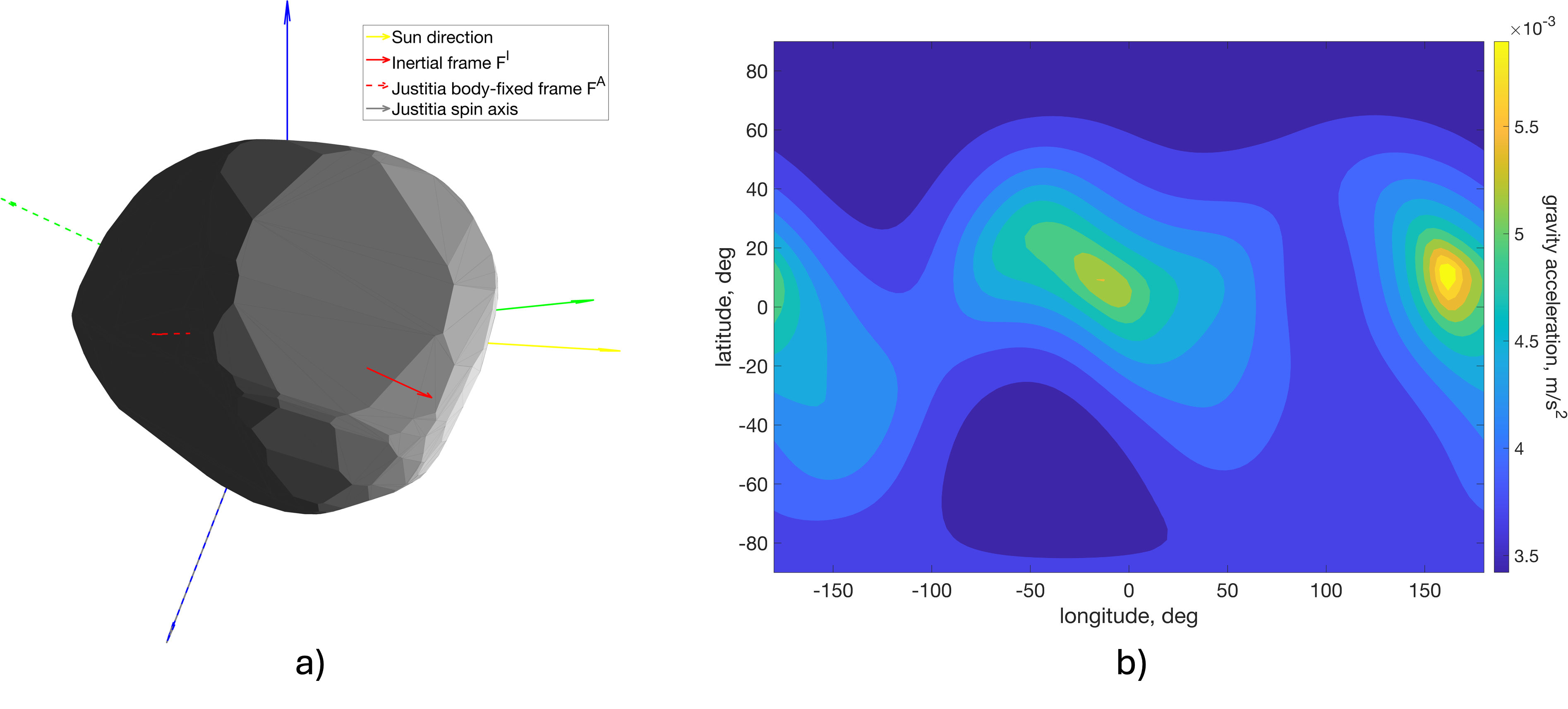}
  \caption{Asteroid Justitia a) Polyhedron Shape Model, b) Gravity acceleration on Brillouin sphere, Mid $\mu$ case}
  \label{fig:shape model and gravity field}
   \end{center}
\end{figure}

Justitia shape model is represented as a polyhedron with triangular faces (see Fig.~\ref{fig:shape model and gravity field},a). The model is comprised of a list of vertices position vectors $\mathbf{r}^{\mathcal{A}}_{\mathrm{V}_i}$ and links between the vertices. The model includes 1144 facets and 574 vertices. The vertices are given in asteroid body-fixed frame $\mathcal{F^A}$ and initially are normalized such that the polyhedron volume is unit~\cite{damit}. The scaling coefficient $s$ is defined as follows
\begin{equation}
   s = \left(\frac{4\pi}{3} \right)^{1/3}R_{\mathrm{eq}}.
\end{equation} 

Justitia is supposed to rotate around its z-axis – primary axis with the maximum inertia moment, with period of $P$ = 33.1 h (classifying it as a slow rotator) and right-handed spin axis $\lambda$ = $73^{\circ}$, $\theta = -81^{\circ} $ given in J2000 ecliptic frame $\mathcal{F^I}$.

Accurate representation of the gravitational field in the vicinity of an irregular asteroid is crucial for proximity operations, and descent trajectory propagation. Traditional spherical harmonic models lose convergence near the surface due to the irregular geometry of small bodies (i.e., within the Brillouin sphere). Therefore, we adopt the polyhedron gravity model originally formulated by Werner \& Scheeres in~\cite{werner1996exterior}, which yields exact closed-form expressions for the potential and acceleration of a constant-density polyhedral body.
The polyhedron gravity model is adopted for the present mission design and analysis. However, it assumes a uniform bulk density for the asteroid and thus does not consider local mass anomalies. After in-situ gravity characterization, the gravity field can be represented using a spherical-harmonic expansion outside the Brillouin sphere, while the interior (near-surface) field can be treated with the interior spherical Bessel / harmonic framework to avoid divergence issues and improve convergence in close-proximity operations ~\cite{takahashi2014small}. Alternative approaches for interior gravity field modeling of irregular small bodies have also been proposed in~\cite{fereoli2026interior}.

In our work, we use the open-source GaMa Toolkit~\cite{GamaToolkit} implementation of the Werner–Scheeres polyhedron model, which has been validated for small-body environments and is suitable for trajectory analysis close to the asteroid surface. Figure~\ref{fig:shape model and gravity field},b depicts asteroid attraction on the Brillouin sphere ($r_{\mathrm{B}} = \max(r_{\mathrm{V}_i}) = 38.4~\mathrm{km}$) for the Mid $\mu$ Justitia case. 

The Cartesian position and velocity of asteroid (269) Justitia used in this study are obtained from the JPL Small-Body Database ephemeris distributed in SPICE SPK format, accessed via the NASA JPL Horizons system~\cite{JPL_Horizons}. The SPICE kernel is loaded and processed using the NAIF MICE toolkit for MATLAB~\cite{NAIF_SPICE}.

\section{EMA Probe Specifications and Mission ConOps}
\label{sec: Lander description}

The section provides an overview of the EMA proximity operations and outlines the ConOps for the EMA probe mission. Based on the high-level mission requirements and probe technical specifications it introduces the probe attitude control approach. The mission design is largely determined by the initial design constraints, which are defined by the following mission requirements.

Probe design constraints:
\begin{enumerate}
    \item Probe mass is limited to 9.9 kg while dimensions should fit to the cylinder of 320 mm diameter and 300 mm height.
    \item The probe shall not be equipped with a propulsion system or any deployable structures.
    \item The probe shall be powered exclusively by a battery and shall not include any recharging devices, such as solar panels.
    \item The probe shall be equipped with a one-way communication link for transmitting data to the main S/C.
\end{enumerate}

The requirements for the deployment orbit and the probe separation conditions from the main S/C are as follows:

\begin{enumerate}
    \item The periapsis altitude $h_p$ of the main spacecraft shall be at least 5 km above the asteroid surface ($h_p \geq 5$~km). Because the asteroid has a complex shape, the altitude is currently measured relative to the equivalent sphere radius, i.e., $r_p =R_{eq} + h_p$.
    \item The orbit of the main spacecraft shall remain outside the asteroid shadow.
    \item Deployment shall occur at apoapsis in anti-tangential direction using a spring-based separation mechanism that provides a nominal separation velocity of $\Delta V = 1.243$ m/s for the considered probe mass. The current estimate of the separation-velocity magnitude uncertainty is $\pm 10\%$ of the nominal value.
    \item The deployment mechanism may generate a disturbance torque resulting in a maximum initial angular velocity of 2 deg/s about each probe body-fixed axis.
    \item The probe descent trajectory shall ensure touchdown on the asteroid surface during the first revolution following separation from the main spacecraft.
\end{enumerate}

Another major mission design driver is the uncertainty in the asteroid density and internal mass distribution, and consequently in its gravity field. Thus, at the current stage of the mission development, it is only possible to select preliminary reference trajectory and finalize it upon rendezvous with the asteroid and its detailed surveillance. 

The requirements are subsequently used to derive the mission concept, develop the deployment orbit selection strategy, select the reference deployment orbits, and analyze the mission scenarios by evaluating the asteroid surface coverage achieved along the reference trajectories under the considered attitude modes.

\subsection{Probe Technical Specifications}

The EMA probe is a nanosatellite developed in accordance with the mission requirements and is equipped with two payload cameras for asteroid observation and an ADCS for payload pointing, in addition to the necessary satellite subsystems. The probe is designed to operate as a fully autonomous system, as no real-time commanding from the main S/C is available either prior to deployment during the cruise phase of the mission or after deployment during the descent phase.

Let us consider the engineering axes to simulate attitude dynamics as the probe's geometry is defined in the axes, meaning the unit direction vectors for the sensors, payload, and other components. Also, deployment conditions as well as the target attitude are initially defined with respect to the engineering axes.

\begin{figure}[h]
 \begin{center}
  \includegraphics[width=0.5\textwidth]{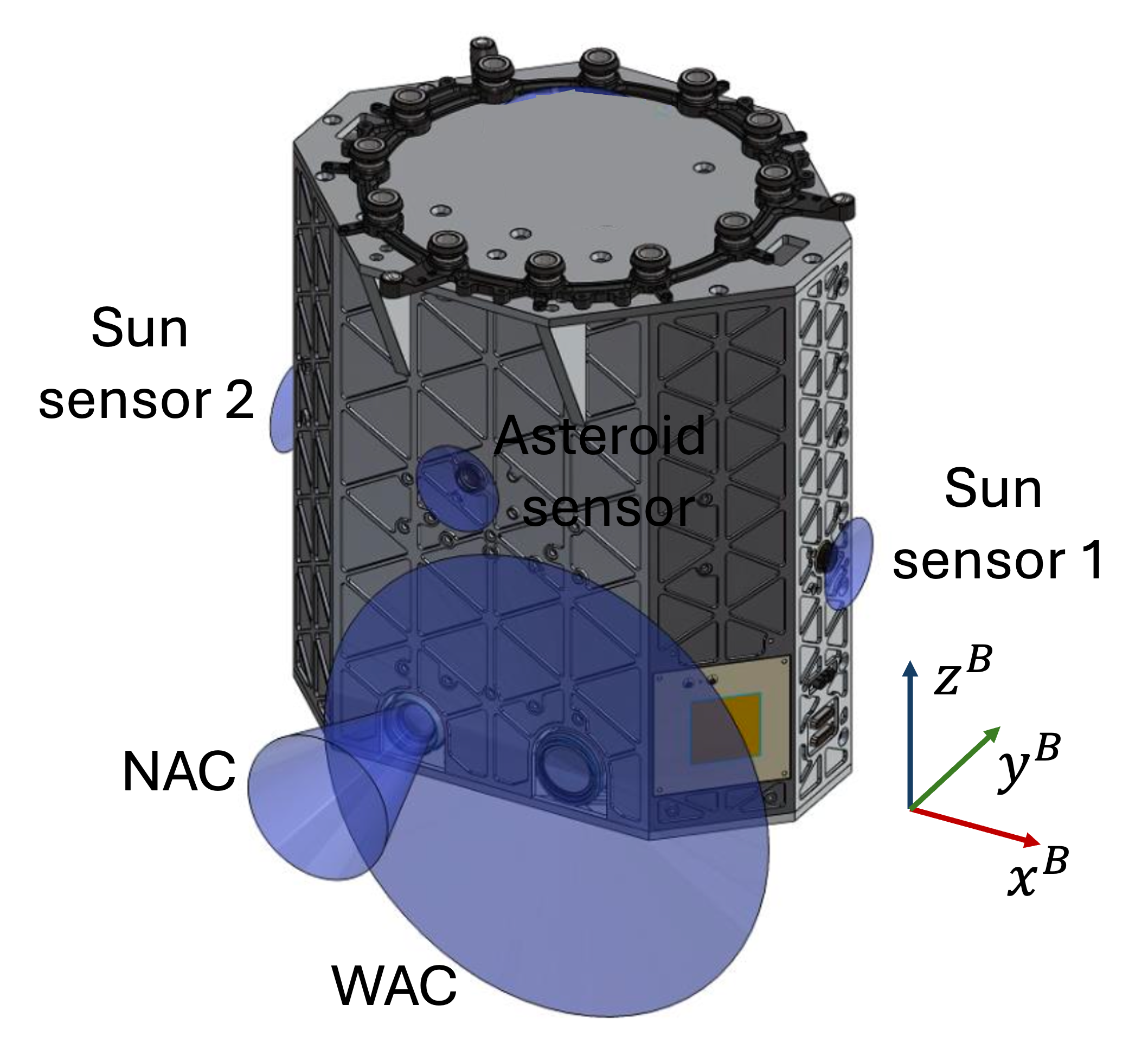}
  \caption{Probe geometry and body-fixed reference frame $\mathcal{F^B}$ with the sensors and payload cameras field-of-view cones}
  \label{fig:probe_geometry}
   \end{center}
\end{figure}

Probe has a mass of 9.87 kg and the inertia tensor given with respect to the body-fixed frame $\mathcal{F^B}$ as follows:

\begin{equation}\label{eq:J_B}
 \mathbf{J} =   
    \begin{bmatrix}
       0.8728 & -0.0009 & 0.0030 \\
       -0.0009 & 0.8946 & 0.0007 \\
       0.0030 & 0.0007 & 0.0566
    \end{bmatrix} \mathrm{kg \, m^2}.
\end{equation}

Figure~\ref{fig:probe_geometry} shows the probe geometry, the adopted body-fixed reference frame $\mathcal{F^B}$, and the FOV cones of the sensors and payload cameras. Based on this frame, the unit direction vectors of the payload cameras, Sun sensors, asteroid sensor, and antennas are defined as follows:
\begin{subequations}
\begin{align}\label{eq: direction vectors}
\hat{\mathbf{r}}_{\mathrm{NAC}}^{\mathcal{B}} &= [0, -1, 0]^{\top}, \\
\hat{\mathbf{r}}_{\mathrm{WAC}}^{\mathcal{B}} &= [0, -1, 0]^{\top}, \\
\hat{\mathbf{r}}_{\mathrm{SS}_1}^{\mathcal{B}} &= [1, 0, 0]^{\top}, \\
\hat{\mathbf{r}}_{\mathrm{SS}_2}^{\mathcal{B}} &= [-1, 0, 0]^{\top}, \\
\hat{\mathbf{r}}_{\mathrm{AS}}^{\mathcal{B}} &= [0, -1, 0]^{\top}, \\
\hat{\mathbf{r}}_{\mathrm{A}_1}^{\mathcal{B}} &= [\frac{1}{\sqrt{2}},\frac{1}{\sqrt{2}}, 0]^{\top}, \\
\hat{\mathbf{r}}_{\mathrm{A}_2}^{\mathcal{B}} &= [-\frac{1}{\sqrt{2}},-\frac{1}{\sqrt{2}}, 0]^{\top},
\end{align}
\end{subequations}

\noindent where $\mathrm{NAC}$ stands for narrow angle camera, $\mathrm{WAC}$ stands for wide angle camera, $\mathrm{SS}_1$, $\mathrm{SS}_2$ are for the first and second Sun sensors, $\mathrm{AS}$ stands for the asteroid sensor, $\mathrm{A}_1$, $\mathrm{A}_2$ stand for the antennas.

The probe attitude determination and control system (ADCS) employs four reaction wheels (RWs) arranged in a regular tetrahedral configuration.
The corresponding reaction wheel configuration matrix 
$\mathbf{A}_{\mathrm{rw}}$, whose columns represent the unit spin-axis vectors expressed in the body frame $\mathcal{F^B}$, is given by
\begin{equation}\label{eq:A_rw}
\mathbf{A}_{\mathrm{rw}}=
\begin{bmatrix}
\;\;0.892586 & 0        & -0.892586 & 0 \\
\;\;0        & 0.892586 & 0         & -0.892586 \\
\;\;0.450878 & 0.450878 & 0.450878  & 0.450878
\end{bmatrix}.    
\end{equation}

Each RW has a maximum torque of $7~\mathrm{mNm}$, a maximum angular velocity of $10{,}000~\mathrm{rpm}$, a moment of inertia about its symmetry axis of $J_{\mathrm{rw}} = 2.6\times10^{-5}~\mathrm{kg \, m^2}$, and a mass of $0.144~\mathrm{kg}$. The reaction wheels are sized such that they do not saturate during the $\sim2.5$ hours descent phase.

The probe payload includes two cameras, WAC and NAC, with half FOV of $\gamma_{\mathrm{WAC}} = 40^{\circ}$ and $\gamma_{\mathrm{NAC}} = 7^{\circ}$ and angular resolutions of $\gamma_{\mathrm{WAC}}^{\mathrm{res}} = 2.12'$ and $\gamma_{\mathrm{NAC}}^{\mathrm{res}} = 15.7''$, respectively.

\subsection{Mission Scenario and Concept of Operations}

\begin{figure}[h]
 \begin{center}
  \includegraphics[width=1\textwidth]{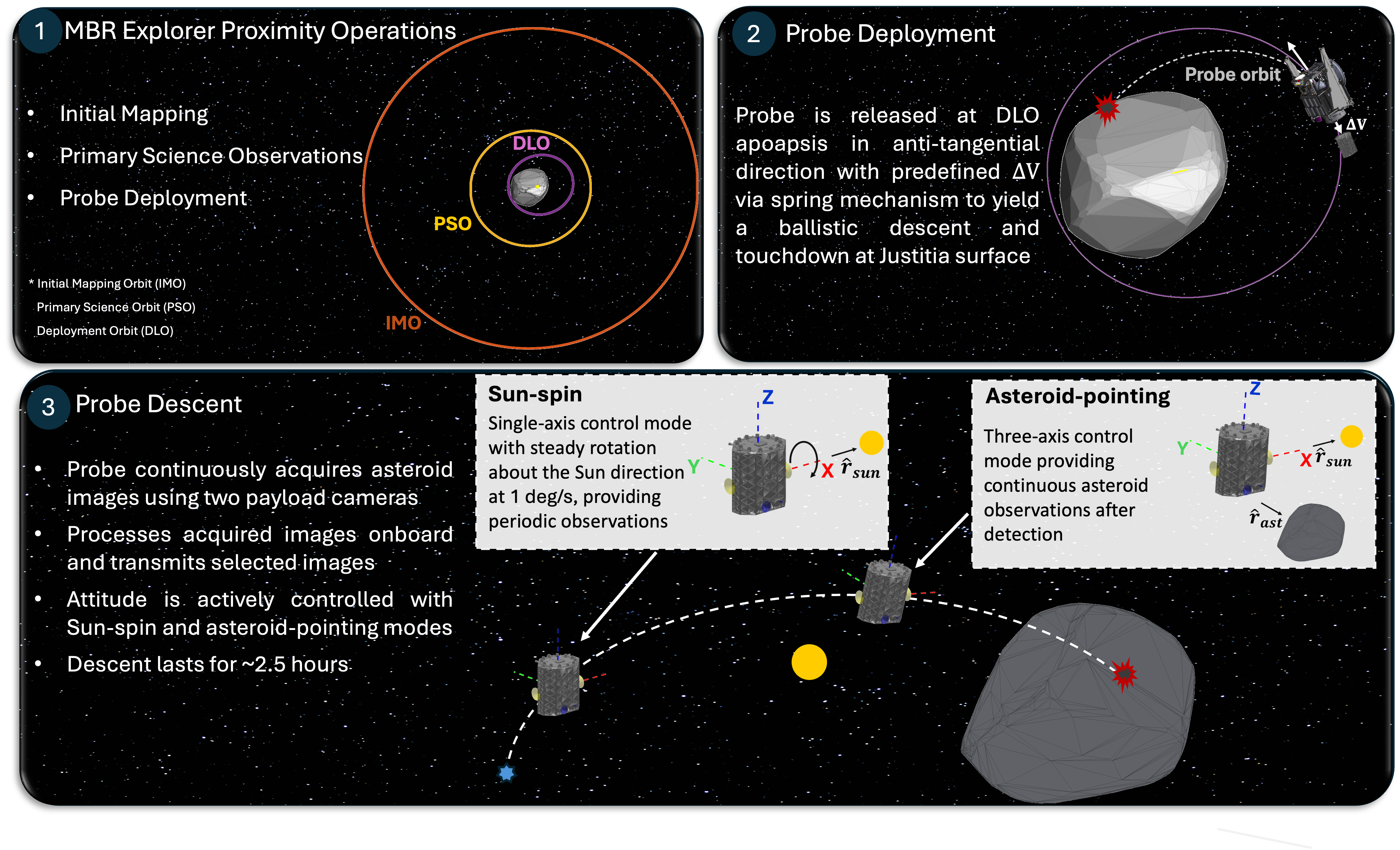}
  \caption{EMA Probe Mission ConOps}
  \label{fig:EMA Probe ConOps}
   \end{center}
\end{figure}

After seven years of interplanetary flight, including three gravity-assist maneuvers and six asteroid flybys, the main spacecraft will rendezvous with asteroid Justitia in 2034 to conduct the proximity operations campaign and deploy the EMA probe. Figure~\ref{fig:EMA Probe ConOps} illustrates the high-level ConOps for the EMA probe mission. The proximity operations of the main S/C are planned to last approximately six months which include three key phases: initial asteroid mapping, primary science observations, and probe deployment~\cite{parkerproximity}. During these operations, the detailed surface topography and gravity field of the asteroid will be reconstructed, and its physical properties and composition will be investigated using the main spacecraft's payload instruments~\cite{almazmi2023scienceB}. Finally, after completing its primary scientific campaign, the main spacecraft will transfer to the deployment orbit (DLO) where the probe will be deployed.

The probe is to be released from the main S/C at the apoapsis of an elliptical orbit around Justitia using a spring deployment mechanism with a predefined release $\Delta V$ for the given probe mass. At deployment, the probe attitude is aligned with the orbital reference frame, also called the Hill frame, with its body-fixed $z$-axis directed along the orbital velocity vector and its body-fixed $-y$-axis, corresponding to the camera boresight, directed toward the asteroid. Separation is performed in the direction opposite to the orbital velocity vector, thereby lowering the periapsis of the probe descent orbit. The resulting periapsis is required to lie below the asteroid surface to ensure probe touchdown.

After separation, the probe enters its descent operation mode. During descent, it continuously acquires asteroid images using the NAC and WAC, processes the acquired imagery using the onboard digital processing unit~\cite{Sirota2025}, ranks the images according to the predefined selection criteria, and transmits the highest-ranked images to the main S/C. In parallel, the ADCS controls the probe attitude to maintain the required camera pointing toward the asteroid. The onboard image-selection algorithm assigns a quality score to each acquired image using a weighted combination of image quality metrics related to exposure, contrast, structural content, and sharpness. Images are acquired at 2 Hz, and the highest ranked images are selected periodically for transmission. The detailed scoring formulation and onboard implementation are currently under development and are outside the scope of this paper.

Active payload camera pointing is an important feature of the probe because the deployment mechanism may generate a disturbance torque and induce probe tumbling. In addition, the exact deployment orbit will be finalized only during proximity operations at Justitia and cannot be uploaded to the probe due to the one-way communication architecture. Therefore, the attitude determination and control approach shall be independent of prior knowledge of the exact deployment orbit and remain applicable to any orbit satisfying the prescribed deployment requirements. The approach utilizes the required near-terminator orientation of the deployment orbit. Two reference directions are used for probe attitude determination and control: the Sun direction $\hat{\mathbf{r}}_{\mathrm{Sun}}$ and the asteroid direction $\hat{\mathbf{r}}_{\mathrm{ast}}$. These directions are measured using a pair of fine Sun sensors and an asteroid sensor, respectively. As presented in the previous section, the asteroid sensor and Sun sensors are aligned along perpendicular body-fixed directions.

During descent phase, the probe ADCS operates in two primary control modes: Sun-spin and asteroid-pointing. Sun-spin is the first control mode to run. It de-tumbles the probe, points Sun Sensor 1 toward the Sun, and starts controlled steady rotation around the Sun direction at a rate of 1 deg/s. As the deployment orbit is near-terminator, the rotation of the probe around the Sun direction, with its $+x$ side pointing toward the Sun, ensures that the asteroid is visible to the asteroid sensor aligned with the $-y$ direction. Once asteroid is detected by the asteroid sensor, probe's ADCS transfers to the asteroid-pointing mode which performs three-axis attitude control with the target orientation defined by the Sun and asteroid direction vectors. If, for any reason, the asteroid is not detected, the probe remains in the Sun-spin mode. In this mode, continuous rotation about the Sun direction provides periodic asteroid observation opportunities throughout the descent, which is considered acceptable for the mission. Therefore, the Sun-spin mode also serves as a safe mode for the probe.

The target attitude for the asteroid-pointing mode is defined as follows. The probe’s body-fixed $+x$-axis is pointed to the Sun. Therefore camera boresight direction or body-fixed $-y$-axis lies in the terminator plane. Its exact direction is set by a rotation angle around $+x$ axis. The goal is to have the asteroid as close to the boresight of the camera as possible, which is when asteroid center is located on the horizontal axis of the camera. In terms of geometry, the $-y$ projection to the asteroid direction $\hat{\mathbf{r}}_{\mathrm{ast}}$ is the biggest.

The target asteroid-pointing attitude $\mathbf{C}^{\mathcal{IB}_{\mathrm{req}}} = \{\hat{\mathbf{x}},\hat{\mathbf{y}},\hat{\mathbf{z}}\}$ is defined as follows:

\begin{equation}\label{eq:asteroid pointing attitude}
\hat{\mathbf{x}} = \hat{\mathbf{r}}_{\mathrm{Sun}}^{\mathcal{I}}, \quad \hat{\mathbf{z}} = \frac{\hat{\mathbf{r}}_{\mathrm{ast}}^{\mathcal{I}} \times \hat{\mathbf{r}}_{\mathrm{Sun}}^{\mathcal{I}}}{||\hat{\mathbf{r}}_{\mathrm{ast}}^{\mathcal{I}} \times \hat{\mathbf{r}}_{\mathrm{Sun}}^{\mathcal{I}}||}, \quad \hat{\mathbf{y}} = \hat{\mathbf{z}} \times \hat{\mathbf{x}},
\end{equation}

\noindent let us note that $\hat{\mathbf{r}}_{\mathrm{ast}} \times \hat{\mathbf{r}}_{\mathrm{Sun}} \neq 0$ throughout descent, or in other words phase angle $\alpha$ is never equal to zero or 180 deg. Therefore the target attitude is always defined. It is also important to highlight that within the attitude determination algorithm $\hat{\mathbf{r}}_{\mathrm{ast}}$ will be obtained from the asteroid sensor measurement and thus can have a deviation from the ideal one.

The descent orbit does not necessarily coincide exactly with the terminator plane and will generally have a certain angular offset from it. The proposed attitude control logic relies on the Sun direction being readily detectable using two wide FOV Sun sensors and on the asteroid entering the asteroid-sensor FOV as the probe rotates about the Sun direction. Therefore, the deployment orbit must be selected such that, when the target attitude defined by Eq.~\eqref{eq:asteroid pointing attitude} is maintained, the payload cameras retain sufficient capability to observe the asteroid. This requirement is particularly important for the NAC because of its limited FOV.

The preliminary control algorithms utilizing Lyapunov function based feedback control laws are outlined in~\ref{app: attitude control algorithms}. Detailed analysis including sensor errors, actuator uncertainties, controller gain optimization, and Monte Carlo dispersion assessment is outside the scope of the paper.

\section{Deployment Orbit Selection}
\label{sec:deployment orbit selection}

The section describes the deployment orbit selection, where the probe will be released from the main S/C. As the probe is not equipped with a propulsion system, the only opportunity for the probe mission design team to satisfy the mission objectives is to select a feasible orbit to be released from. The orbit selection strategy utilizes knowledge on the main S/C proximity operation orbits analysis from~\cite{parkerproximity} and extends the analysis presented in~\cite{ema_biktimirov}. While recent investigations have examined the general dynamical properties and orbital stability regions around Justitia~\cite{Braga_2026}, the present study focuses specifically on deployment orbit geometry driven by illumination and imaging constraints and taking into account the attitude control approach.

The section derives the method for consecutive deployment orbit selection starting with formal requirements for the probe deployment orbit, selecting the orbit starting from the orbit ellipse consideration, then moving to the orbit attitude selection approach - introducing the terminator frame method and camera boresight deviation criterion, and finally completes the chapter with presenting and analyzing example reference orbits.

The formal requirements for the probe deployment orbit were derived through an iterative mission design process and constitute one of the outcomes of the present analysis. They enable the selection of a deployment orbit that remains feasible under uncertainties in the asteroid parameters while supporting the proposed attitude determination and control approach throughout the descent. The requirements are as follows:

\begin{enumerate}
     \item Probe's descent trajectory shall pass over the lit part of Justitia, while having phase angle $\alpha(t) \leq \pi/2$ throughout the descent
     
    \item For a given main S/C orbit periapsis $r_{\mathrm{p}}$, the deployment orbit apoapsis $r_{\mathrm{a}}$ should be chosen to yield a probe time of flight (TOF) of $T_{\mathrm{desc}} = 2.5$ hours, with an acceptable descent-duration tolerance of $\pm 30$~min
    
    \item Probe's primary attitude control mode is asteroid pointing, at which its $+x$ axis points to the Sun direction, while probe's $-y$ axis should have maximum possible $-y$ axis projection to the asteroid direction vector
    
    \item The deployment orbit attitude shall be selected in a way allowing to perform the asteroid observation by both probe cameras, NAC and WAC, throughout the entire descent, i.e. for the target attitude of the asteroid pointing mode, the asteroid shall be within the FOVs of the NAC and WAC throughout the descent
\end{enumerate}

The deployment orbit selection is performed in two consecutive steps. First, the orbit size and shape are determined by selecting the semi-major axis and eccentricity required to provide the prescribed descent duration. Second, the orbit attitude is selected by determining the inclination, RAAN, and AOP subject to the illumination and camera observation constraints.

\subsection{Orbit Size Selection} 
Let us consider the Keplerian deployment orbit of the main S/C, from which the probe is released onto a ballistic descent trajectory. The main S/C periapsis altitude is fixed at 5 km, while the apoapsis radius is varied. Different apoapsis radii result in different probe descent durations, $T_{\mathrm{desc}}$, and touchdown velocities. As gravity field of Justitia is not well defined yet, we consider different scenarios of gravitational parameter $\mu$ corresponding to uniform asteroid matter densities (see Section~\ref{sec:Justitia properties}).  
Let’s also evaluate duration of the line-of-sight (LOS) presence between the main S/C and probe after its touchdown for the trajectories yielding different descent time. 

\begin{figure}[h!]
 \begin{center}
  \includegraphics[width=0.8\textwidth]{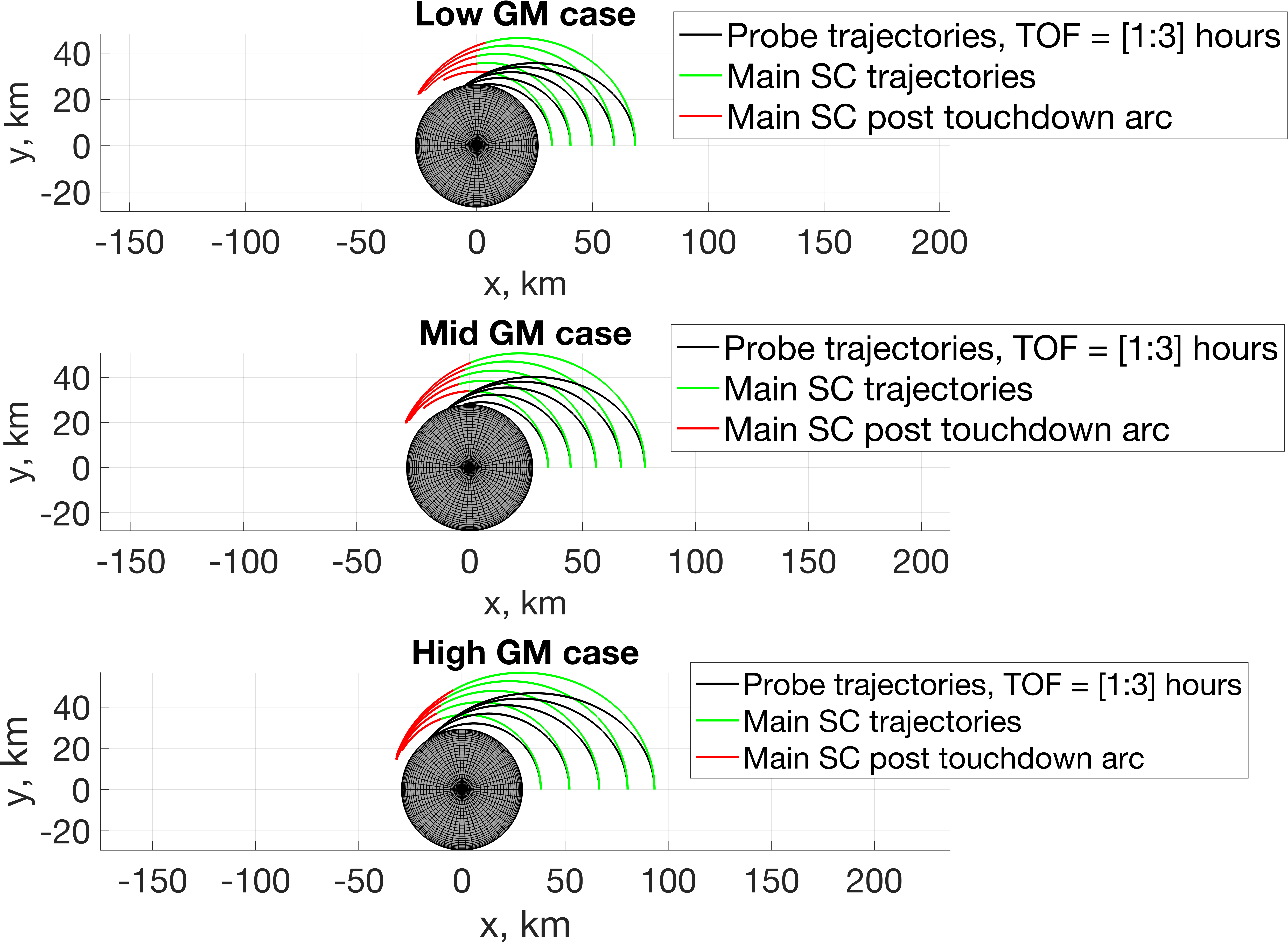}
  \caption{Probe and main spacecraft trajectories after separation at the deployment-orbit apoapsis for the Low-, Mid-, and High-$\mu$ cases. The trajectories correspond to descent times between 1 and 3~h, obtained by varying the apoapsis radius while keeping the periapsis radius fixed}
  \label{fig:descent trajectories}
   \end{center}
\end{figure}

Figure~\ref{fig:descent trajectories} demonstrates the trajectories of the main S/C and probe after separation at the deployment orbit apoapsis for the point mass dynamics. The asteroid shape for the analysis is approximated by a sphere of the asteroid equivalent radius $R_{\mathrm{eq}}$. Thus, the touchdown condition for the probe is to have a position vector length equal to the corresponding equivalent radius $R_{\mathrm{eq}}$.

\begin{figure}[h!]
 \begin{center}
  \includegraphics[width=0.9\textwidth]{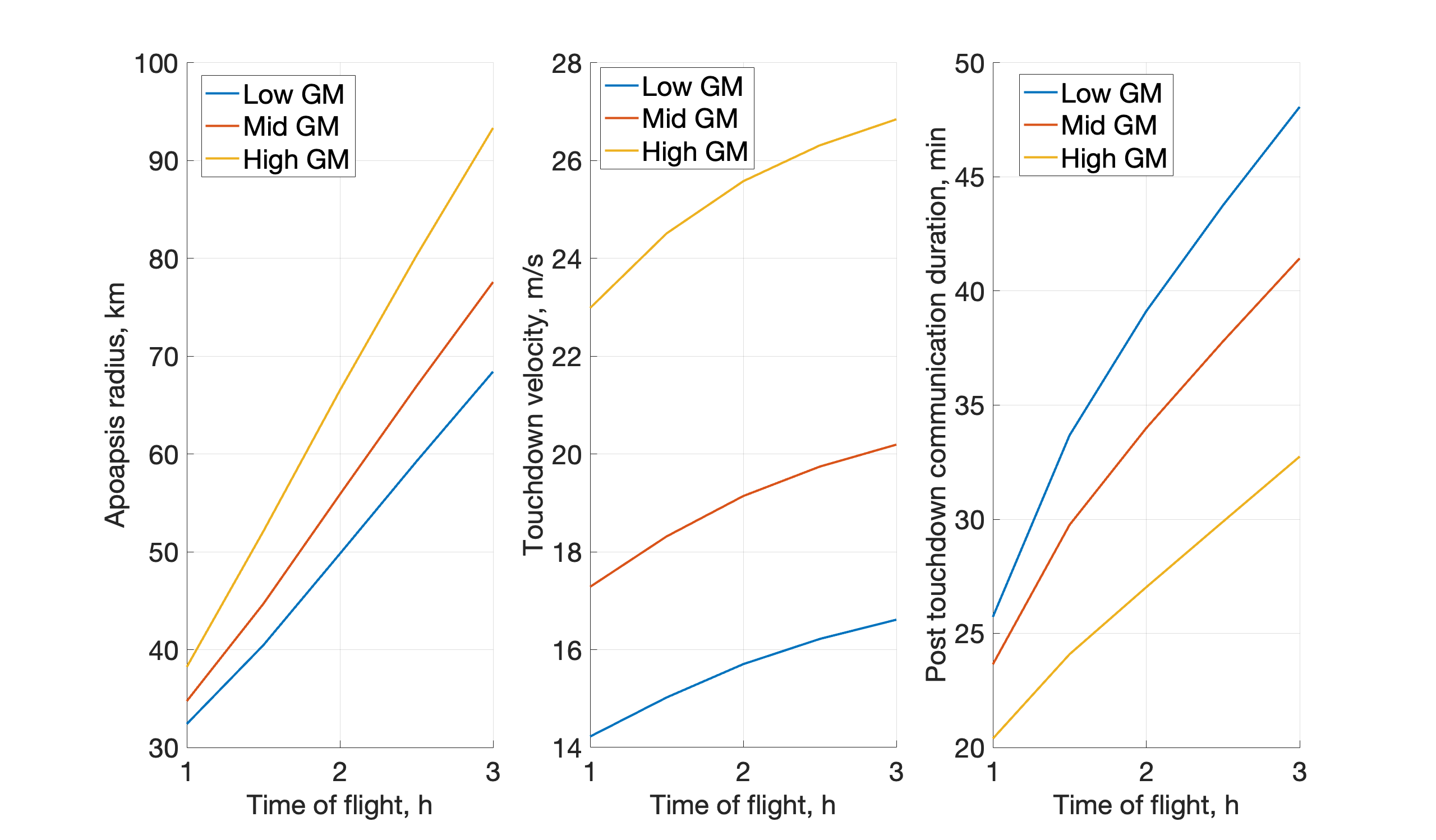}
  \caption{Metrics of the time-constrained descent trajectories shown in Fig.~\ref{fig:descent trajectories}: apoapsis radius, touchdown velocity, and post-touchdown LOS presence duration as functions of descent time for the Low-, Mid-, and High-$\mu$ cases}
  \label{fig:orbit size selection}
   \end{center}
\end{figure}

The trajectories of the main S/C are selected in a way to yield the probe descent time to be in a range from 1 to 3 hours for the given deployment $\Delta V$. The green arcs show portions of the main spacecraft trajectory during which the probe remains in orbit, i.e., does not touch down, while the red arc shows the portion of the main spacecraft trajectory during which the probe completes its descent while there is a LOS between the main spacecraft and the probe. Figure~\ref{fig:orbit size selection} reveals corresponding parameters of the various descent trajectories such as the orbit apoapsis, probe touchdown velocity, duration of LOS presence between probe and main S/C after probe touchdown. 

Let us note that touchdown velocity varies from $\sim$15 m/s to $\sim$30 m/s depending on the considered asteroid gravitational parameter and deployment orbit apoapsis. We can conclude that, for the considered range, there are no conditions yielding a soft touchdown, assuming that a soft touchdown can be performed at a maximum velocity of a few meters per second, whereas in this analysis touchdowns occur at a few tens of meters per second. Therefore, the post touchdown operation is not considered in the study and in the probe mission in general, while according to the post touchdown LOS presence duration with the main S/C, probe can still transmit images for the period of time varying from $\sim 20$ to $\sim 50$ min.

Current baseline solution for the probe deployment orbit is selected to provide a nominal TOF of 2.5 hours to ensure execution of the probe descent program for the considered Justitia gravity scenarios. The nominal TOF is selected to provide sufficient margin within approximately 4 hours of battery-supported probe operation, accounting for the nominal battery degradation during the cruise phase. It also accounts for the acceptable TOF tolerance of \(\pm 30\)~min and leaves additional margin for possible data transmission during the post-touchdown LOS period of approximately 20--50 min, should the probe remain operational after touchdown.

This yields the following pairs of ($r_{\mathrm{a}}$, $r_{\mathrm{p}}$) and ($a$, $e$) for the main S/C (see Table~\ref{tab:orbit sizes, sc}) and probe orbit ellipses (see Table~\ref{tab:orbit sizes, lander}).

\begin{table}[h]
\centering
\caption{Keplerian S/C orbit sizes for different Justitia $\mu$ cases}
\label{tab:orbit sizes, sc}

\begin{tabular}{lll}
\toprule
Case & ($r_\mathrm{a}$ [km], $r_\mathrm{p}$ [km]) & ($a$ [km], $e$) \\
\midrule

Low $\mu$  & $(59.2, 31.6)$ & $(45.4, 0.3043)$ \\
Mid $\mu$  & $(66.9, 33)$   & $(49.97, 0.3396)$ \\
High $\mu$ & $(80.3, 34.4)$ & $(57.3, 0.4)$ \\

\bottomrule
\end{tabular}

\end{table}

\begin{table}[h!]
\centering
\caption{Keplerian probe orbit sizes for different Justitia $\mu$ cases}
\label{tab:orbit sizes, lander}

\begin{tabular}{lll}
\toprule
Case & ($r_{\mathrm{a}}$ {[}km{]}, $r_{\mathrm{p}}$ {[}km{]}) & (a {[}km{]}, e) \\
\midrule

Low $\mu$  & (59.2, 19.4) & (39.3, 0.5065) \\
Mid $\mu$  & (66.9, 21.6) & (44.3, 0.5117) \\
High $\mu$ & (80.3, 24.2) & (52.3, 0.5363) \\

\bottomrule
\end{tabular}

\end{table}

\subsection{Orbit Attitude Selection -- Terminator Frame Method} 
\begin{figure}[h!]
 \begin{center}
  \includegraphics[width=0.8\textwidth]{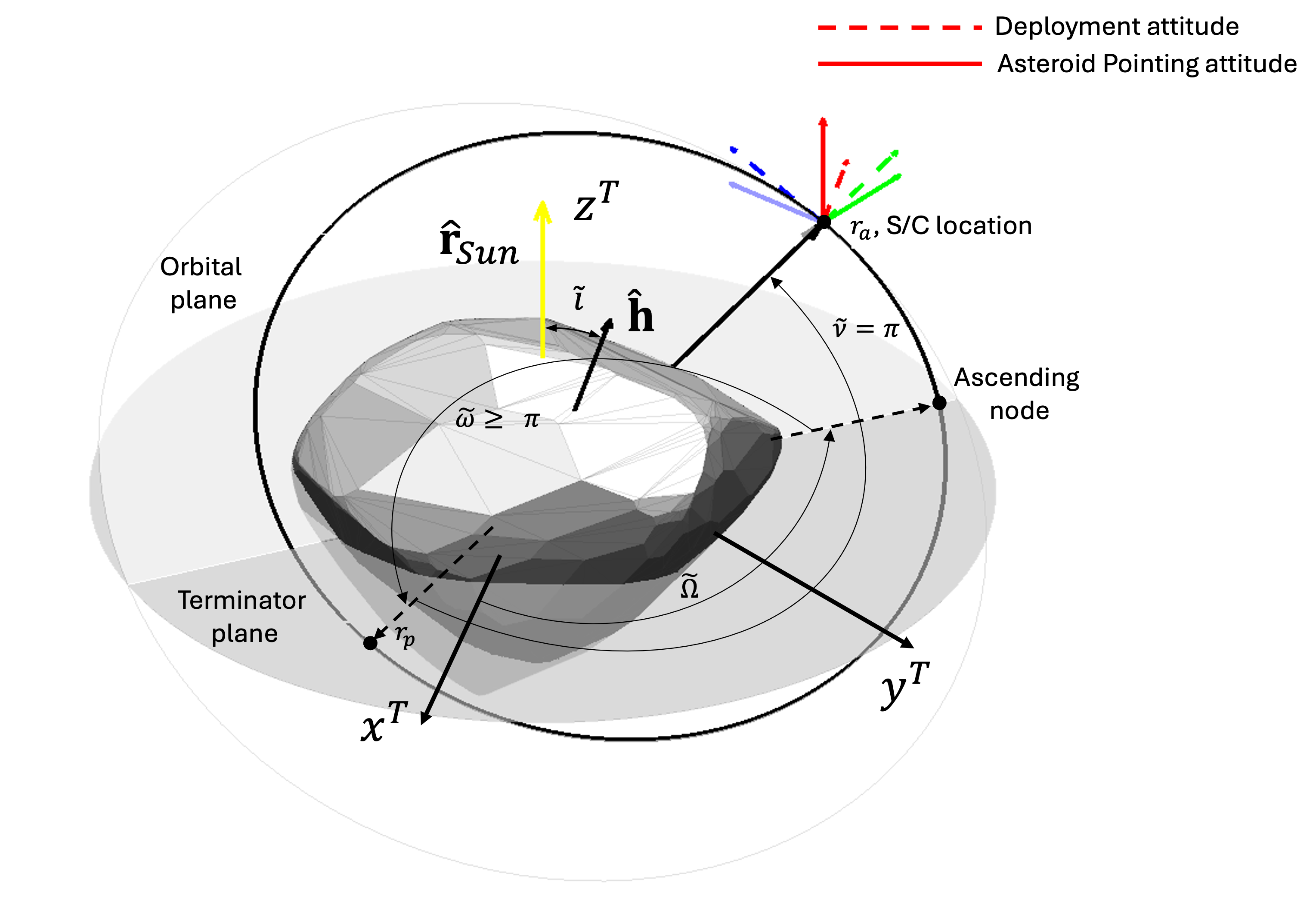}
  \caption{Terminator plane angle selection geometry}
  \label{fig:terminator angle selection}
   \end{center}
\end{figure}

After selecting the deployment orbit ellipse, the deployment orbit attitude is considered. A Keplerian orbit attitude is defined by a set of three Euler angles denoted as $i, \Omega, \omega$ - inclination, RAAN, and AOP. Let us introduce an auxiliary reference frame, denoted as a terminator frame $\mathcal{F^T}$ (see Fig.~\ref{fig:terminator angle selection}) for the purpose of the deployment orbit selection formal procedure.

The reference plane of the terminator frame will be the terminator plane, while the reference frame basis is as in Eq.~\ref{eq:terminator frame}. The orbital elements expressed in the terminator frame $\mathcal{F^T}$ are denoted by $\widetilde{\textbf{\oe}} = [\tilde{a}, \tilde{e}, \tilde{i}, \tilde{\Omega}, \tilde{\omega}, \tilde{\nu}]$. The semi-major axis $\tilde{a}$ and eccentricity $\tilde{e}$ will be the same as selected previously (see Table~\ref{tab:orbit sizes, lander}) as it does not depend on the reference frame choice, while the other elements will be different depending on the reference frame. Following the requirements list, we can note that the true anomaly of the deployment location on orbit corresponds to the orbit apoapsis, i.e., deployment location's true anomaly $\tilde{\nu} = \pi$. For the problem of the deployment orbit selection under the lighting constraints, $\tilde{i}$ and $\tilde{\omega}$ are coupled. As the probe touchdown should be made within the first revolution after deployment, the probe descent trajectory arc will be less than a half of the orbit (see Fig.~\ref{fig:descent trajectories}). Due to the fact that the deployment will happen at the apoapsis while the probe shall pass over the lit part of Justitia during its descent, we can conclude that the apoapsis shall be close to the ascending node of the orbit represented in the terminator frame $\mathcal{F^T}$. Therefore, argument of periapsis $\tilde{\omega}$ shall follow the condition

\begin{equation}
\tilde{\omega} \geq \pi.    
\end{equation}

The illumination geometry is characterized by the phase angle $\alpha$, defined at the asteroid center as the angle between the Sun direction and the probe position vector:
\begin{equation}
\alpha(t)
=
\arccos
\left(
\hat{\mathbf r}_{\mathrm{Sun}}(t)
\cdot
\hat{\mathbf r}(t)
\right).
\label{eq:phaseAngle}
\end{equation}

Let us note that when $\tilde{\omega} = \pi$ and $\tilde{\nu} = \pi$, the initial phase angle $\alpha_0 = \pi/2$, while we can move the probe deployment point closer to the lit part, i.e. decreasing the initial phase angle $\alpha_0$, by increasing further the AOP $\tilde{\omega}$. Nevertheless, the phase angle change throughout the descent $\alpha(t)$ will depend on the orbit inclination $\tilde{i}$. For example, if the orbit inclination $\tilde{i}$ is zero the phase angle $\alpha$ will remain $\pi/2$ throughout the entire descent. Therefore, we conclude that the inclination shall be positive, $\tilde{i} > 0$, while not being too large to allow the cameras to observe the asteroid during the asteroid-pointing mode. RAAN, $\tilde{\Omega}$, remains a variable parameter that can be adjusted without affecting the selection criteria. It can be used to define the locations on the asteroid over which the probe orbit passes.

\subsubsection{Camera Boresight Deviation Criterion}
We next determine the feasible ranges of the inclination $\tilde{i}$ and argument of periapsis $\tilde{\omega}$ while taking into account the target probe attitude required for asteroid observation. The reference attitude used in the asteroid-pointing mode is defined in Eq.~\ref{eq:asteroid pointing attitude}. Based on this attitude geometry, the camera boresight is aligned with the projection of the asteroid direction onto the plane normal to the Sun direction.

\begin{figure}[h!]
 \begin{center}
  \includegraphics[width=0.4\textwidth]{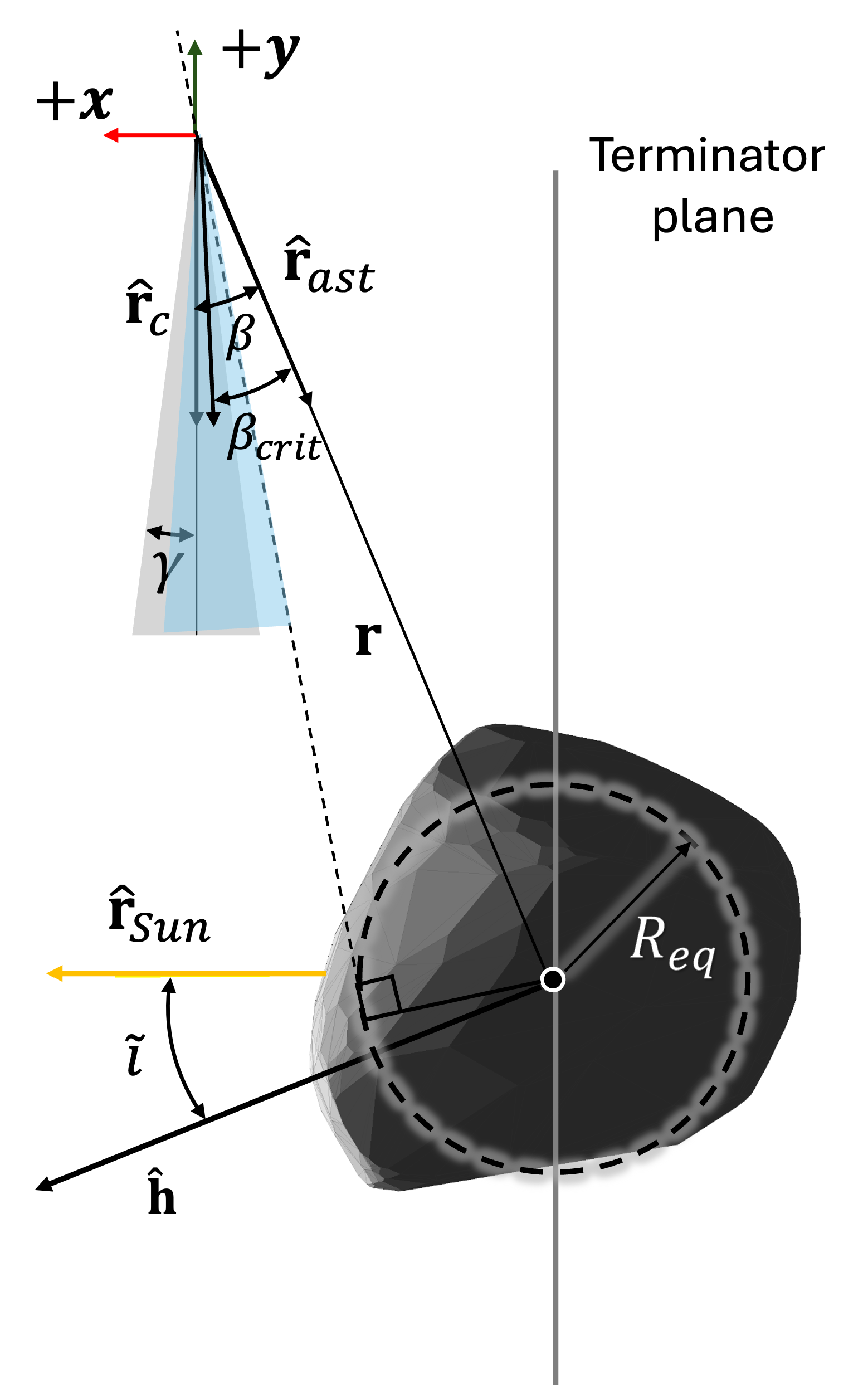}
  \caption{Camera boresight deviation geometry for asteroid observation under a spherical-asteroid approximation. The blue FOV cone represents the critical visibility condition, in which its boundary is tangent to the asteroid surface, whereas the gray cone represents an arbitrary current camera pointing. Both cone axes are shown in the plane containing the probe and the asteroid center.}
  \label{fig:critical observation condition}
   \end{center}
\end{figure}

Figure~\ref{fig:critical observation condition} illustrates the camera pointing geometry used to evaluate asteroid observability. The probe is located at the position
$\mathbf{r}(t)$ relative to the asteroid center, while the camera boresight direction $\hat{\mathbf{r}}_{\mathrm{c}}(t)$ is determined by the reference attitude defined in Eq.~\eqref{eq:asteroid pointing attitude}. The asteroid direction from the probe is denoted by $\hat{\mathbf{r}}_{\mathrm{ast}}(t)=-\hat{\mathbf{r}}(t)$.
For the present analysis, the asteroid is approximated by a sphere of equivalent
radius $R_{\mathrm{eq}}$.

The camera pointing geometry is characterized by the boresight deviation angle $\beta$, the critical deviation angle $\beta_{\mathrm{crit}}$, and the corresponding boresight deviation margin $\Delta\beta$, defined as follows
\begin{align}
\beta(t)
&=
\arccos\left(
\hat{\mathbf{r}}_{\mathrm{c}}(t)
\cdot
\hat{\mathbf{r}}_{\mathrm{ast}}(t)
\right),
\label{eq:beta}
\\
\beta_{\mathrm{crit}}(t)
&=
\arcsin\left(
\frac{R_{\mathrm{eq}}}{r(t)}
\right)
+\gamma,
\label{eq:betaCritical}
\\
\Delta\beta(t)
&=
\beta_{\mathrm{crit}}(t)-\beta(t),
\label{eq:deltaBeta}
\end{align}

\noindent where $\gamma$ is the camera FOV half-angle. The critical angle $\beta_{\mathrm{crit}}$ corresponds to the limiting geometry in which the boundary of the camera FOV is tangent to the spherical asteroid surface. Consequently, $\Delta\beta(t)>0$ indicates that at least a part of the asteroid disk remains within the camera FOV, $\Delta\beta(t)=0$ represents the limiting visibility condition, and $\Delta\beta(t)<0$ indicates complete loss of asteroid visibility.

The observation geometry varies along the descent trajectory, while the goal of the orbit selection is to ensure the asteroid observability throughout the entire descent. For the purpose, we consider the minimum boresight deviation margin $\Delta \beta_{\mathrm{min}}$ within the entire descent trajectory under the assumption of ideal attitude tracking to assess the feasibility of an orbit: 

\begin{equation}
\Delta\beta_{\min} =
\min_{t \in [0,\,T_\mathrm{desc}]}
\left(
\beta_{\mathrm{crit}}(t)-\beta(t)
\right).
\end{equation}

The parameter $\Delta\beta_{\mathrm{min}}$ represents the worst-case angular separation between the camera boresight and the limiting visibility angle defined by the apparent asteroid radius. 

The deployment orbit attitude is considered feasible if the probe remains over the illuminated portion of the asteroid and the prescribed minimum observation margin is maintained throughout the descent:

\begin{subequations}
\begin{align}
\alpha &\leq \pi/2 \label{eq:phase angle req}\\
\Delta\beta_{\mathrm{min}} &\geq \Delta\beta_{\mathrm{min}}^{\ast} \label{eq:boresight angle req}
\end{align}
\end{subequations}

\noindent where $\Delta\beta_{\min}^{\ast}$ is the required minimum boresight deviation margin. The limiting case $\Delta\beta_{\min}^{*}=0$ defines the boundary of geometrically feasible deployment orbit attitudes, whereas a positive value introduces an additional margin for orbit and attitude uncertainties.

\begin{figure}[!h]
 \begin{center}
  \includegraphics[width=1\textwidth]{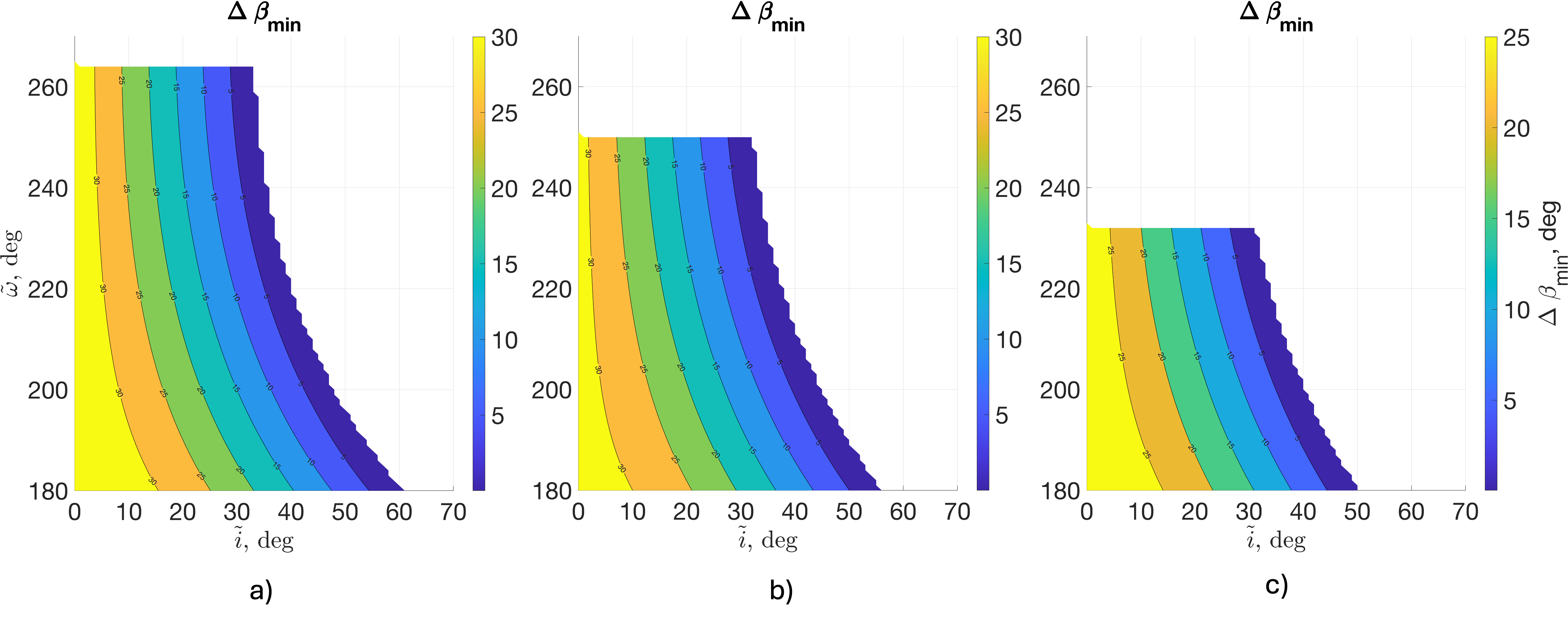}
  \caption{Minimum boresight deviation margin $\Delta\beta_{\min}$ as a function of the terminator frame inclination $\tilde{i}$ and argument of periapsis $\tilde{\omega}$ for the a) Low-$\mu$, b) Mid-$\mu$, and c) High-$\mu$ Justitia cases. Only trajectories satisfying the criteria in Eq.~\ref{eq:phase angle req},\ref{eq:boresight angle req} throughout the descent are shown}
  \label{fig:deltaBetaMinimumMapAllCases}
   \end{center}
\end{figure}

To evaluate feasibility of orbit of different attitudes, a color map of $\Delta \beta_{\mathrm{min}}$ is constructed over a range of terminator frame inclination $\tilde{i}$ and argument of periapsis $\tilde{\omega}$. Figure~\ref{fig:deltaBetaMinimumMapAllCases} shows $\Delta\beta_{\mathrm{min}}$ for the descent trajectories with various inclinations $\tilde{i}$ and AOP $\tilde{\omega}$ where the requirements on the $\Delta \beta_{\mathrm{min}}$ and phase angle $\alpha$ are met considering the boundary case of $\Delta\beta_{\min}^{\ast} = 0$.

Figure~\ref{fig:deltaBetaMinimumMapAllCases} provides a first-order geometrical assessment of the deployment-orbit attitude feasibility. For each considered Justitia gravitational parameter case, the orbit size and eccentricity are fixed according to Table~\ref{tab:orbit sizes, lander}, while the terminator-frame inclination $\tilde{i}$ and argument of periapsis $\tilde{\omega}$ are varied. The resulting maps demonstrate the effect of the deployment orbit attitude on the illumination and camera observation conditions during descent. The contour $\Delta\beta_{\min}=0$ defines the boundary of the geometrically feasible domain, while positive values indicate the available boresight deviation margin. Consequently, a deployment-orbit attitude satisfying any prescribed
positive margin $\Delta\beta_{\min}^{*}$ can be selected directly from the corresponding map.

The figure demonstrates that the range of feasible orbit attitudes depends noticeably on the assumed asteroid size and gravitational parameter. For lower values of $\mu$, the orbit yielding the same descent time has a smaller apoapsis radius, and the asteroid subtends a larger apparent angular size throughout the descent. Consequently, the camera FOV constraint is less restrictive. For higher values of $\mu$, the required descent ellipse becomes larger and more elongated, reducing the apparent angular size of the asteroid and, consequently, the range of deployment orbit attitudes satisfying the camera FOV constraint.

The proposed criterion based on the phase angle and boresight deviation margin allows the deployment orbit attitude selection problem to be treated as a formal feasibility analysis rather than as an unconstrained trial-and-error search or a purely numerical optimization problem without an explicit geometrical interpretation.

\subsection{Reference Deployment Orbits} 

To obtain the Cartesian state of the probe and the main spacecraft at deployment, the orbital elements $\widetilde{\textbf{\oe}}$ defined in the terminator frame are first converted into position and velocity vectors $\mathbf{r}^{\mathcal{T}}$ and $\mathbf{v}^{\mathcal{T}}$. These vectors are subsequently transformed into the inertial frame $\mathcal{F^I}$ using the rotation matrix $\mathbf{C}^{\mathcal{IT}}$ introduced in Eq.~\ref{eq:terminator frame}. 

The deployment epoch plays a critical role in this transformation, as the orientation of the terminator frame $\mathcal{F}^{\mathcal{T}}$ relative to the inertial frame $\mathcal{F^I}$ depends on the instantaneous Sun--asteroid geometry. Therefore, a specific deployment date must be defined to determine the corresponding frame alignment. In this study, the nominal deployment epoch is selected as 1 April 2035, consistent with the mission scenario proposed in~\cite{parkerproximity}.

Let us select candidate orbits for further analysis for all Justitia $\mu$ cases. Following the orbit selection procedure, we consider an orbit with a corresponding descent time of 2.5 hours (see Table~\ref{tab:orbit sizes, lander}). As for the orbit attitude, we shall select the one meeting orbit selection requirements and constraints presented in Eqs.~\ref{eq:phase angle req}, \ref{eq:boresight angle req}. For the purpose of the methodology illustration, let us consider the minimum boresight deviation requirement to be $\Delta\beta_{\mathrm{min}}^{\ast} = \gamma_{\mathrm{NAC}}/2$. Referring to the color maps in Fig.~\ref{fig:deltaBetaMinimumMapAllCases} we can consider $\tilde{i} = 30^{\circ}, \tilde{\omega} = 210^{\circ}$ and $\tilde{\Omega} = 0^{\circ}$ to be within the feasible range. The exact orbits will be selected according to the presented deployment orbit requirement after rendezvous with Justitia and taking into account requirement for the main spacecraft as well. The selected reference orbits and corresponding cartesian states are presented in Tables~\ref{tab:ref_orbit_elements},\ref{tab:ref_cartesian_states}.

\begin{table}[h]
\centering
\caption{Reference descent orbits for different Justitia gravity cases}
\label{tab:ref_orbit_elements}
\resizebox{\linewidth}{!}{%
\begin{tabular}{lcc}
\toprule
Gravity case & $\widetilde{\textbf{\oe}}=[\tilde{a},\,\tilde{e},\,\tilde{i},\,\tilde{\Omega},\,\tilde{\omega},\,\tilde{\nu}]$ & $\textbf{\oe}=[a,\,e,\,i,\,\Omega,\,\omega,\,\nu]$ \\
\midrule
Low $\mu$ &
$[39.3,\ 0.5065,\ 30,\ 0,\ 210,\ 180]$ &
$[39.3,\ 0.5065,\ 64,\ 199.6,\ 30,\ 180]$ \\

Mid $\mu$ &
$[44.3,\ 0.5117,\ 30,\ 0,\ 210,\ 180]$ &
$[44.3,\ 0.5117,\ 64,\ 199.6,\ 30,\ 180]$ \\

High $\mu$ &
$[52.3,\ 0.5363,\ 30,\ 0,\ 210,\ 180]$ &
$[52.3,\ 0.5363,\ 64,\ 199.6,\ 30,\ 180]$ \\
\bottomrule
\end{tabular}

\vspace{1mm}
}
\footnotesize Note: angles are given in degrees; $a$ in km.
\end{table}

\begin{table}[h]
\centering
\caption{Reference descent trajectories Cartesian states at deployment for different Justitia
gravity cases}
\label{tab:ref_cartesian_states}
\resizebox{\linewidth}{!}{%
\begin{tabular}{lll}
\toprule
Gravity case & $\mathbf{X}^{\mathcal{T}} = [\mathbf{r}^{\mathcal{T}}, \mathbf{v}^{\mathcal{T}}]$ & $\mathbf{X}^{\mathcal{I}} = [\mathbf{r}^{\mathcal{I}}, \mathbf{v}^{\mathcal{I}}]$ \\
\midrule

Low $\mu$ &
$[51.31, 25.65, 14.81, -3.32, 4.98, 2.87]$ &
$[44.0, 29.5, -26.6, -4.0, 1.3, -5.2]$ \\

Mid $\mu$ &
$[57.97, 28.99, 16.73, -3.81, 5.72, 3.30]$ &
$[49.7, 33.3, -30.1, -4.57, 1.45, -5.94]$ \\

High $\mu$ &
$[69.53, 34.77, 20.07, -4.52, 6.78, 3.92]$ &
$[59.6, 39.9, -36.1, -5.4, 1.7, -7.0]$ \\

\bottomrule
\end{tabular}

\vspace{1mm}
}
\footnotesize Note: $\mathbf{r}$ in km; $\mathbf{v}$ in m/s.

\end{table}

\begin{figure}[h!]
 \begin{center}
  \includegraphics[width=0.8\textwidth]{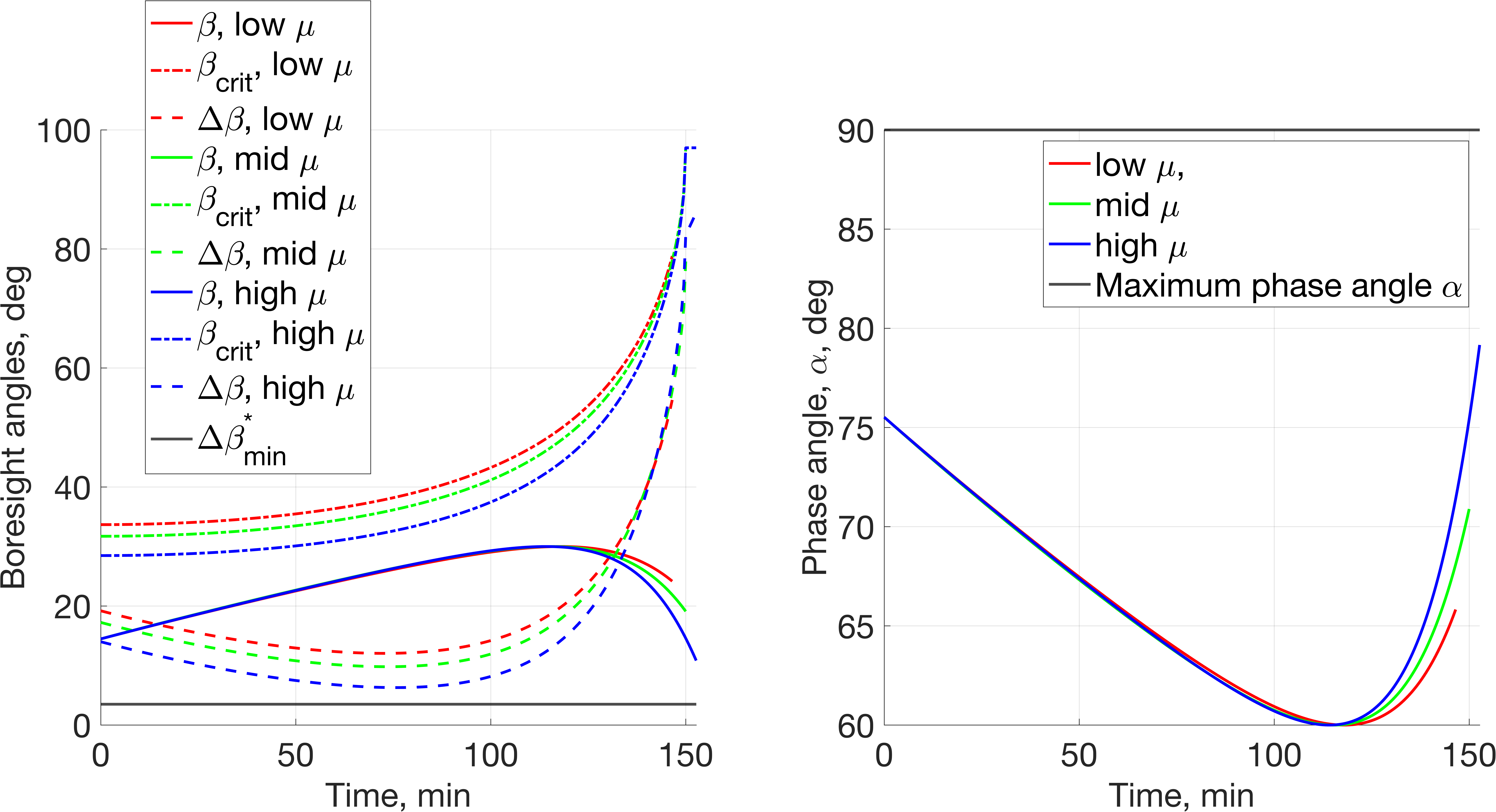}
  \caption{Orbit selection parameters $\alpha$, $\beta$, $\beta_{crit}$, $\Delta\beta $ throughout descent for reference orbits for different Justitia cases}
  \label{fig:mission design parameters}
   \end{center}
\end{figure}

Let us verify the observability of the asteroid within FOV of both cameras for the selected reference orbits. As the required margin $\Delta\beta_{\mathrm{min}}^{\ast}$ is selected for the NAC, it will guarantee the asteroid visibility for the WAC that have a substantially wider FOV. For this purpose, the phase angle $\alpha$ and the boresight deviation angles are calculated throughout the entire trajectories. The left subplot in Fig.~\ref{fig:mission design parameters} demonstrates the boresight deviation angle $\beta(t)$, the critical angle $\beta_{\mathrm{crit}}(t)$, boresight deviation margin and its critical value for all trajectory. It is clear that the trajectories are feasible as the margin $\Delta\beta$ never drops below the prescribed critical value $\Delta \beta_{\mathrm{min}}^{\ast}$. The right subplot shows that the phase angle is always below $90^{\circ}$, which satisfies the orbit requirement.

The point-mass analysis provides a preliminary orbit selection but does not take into account the non-uniform gravity and shape effects encountered near the surface. Therefore, the selected reference trajectories are also propagated using the uniform-density polyhedron model for different initial asteroid rotational phases. The analysis is used to verify the descent duration, touchdown velocity, touchdown location, and first revolution surface intersection condition.

\begin{figure}[h!]
 \begin{center}
  \includegraphics[width=0.7\textwidth]{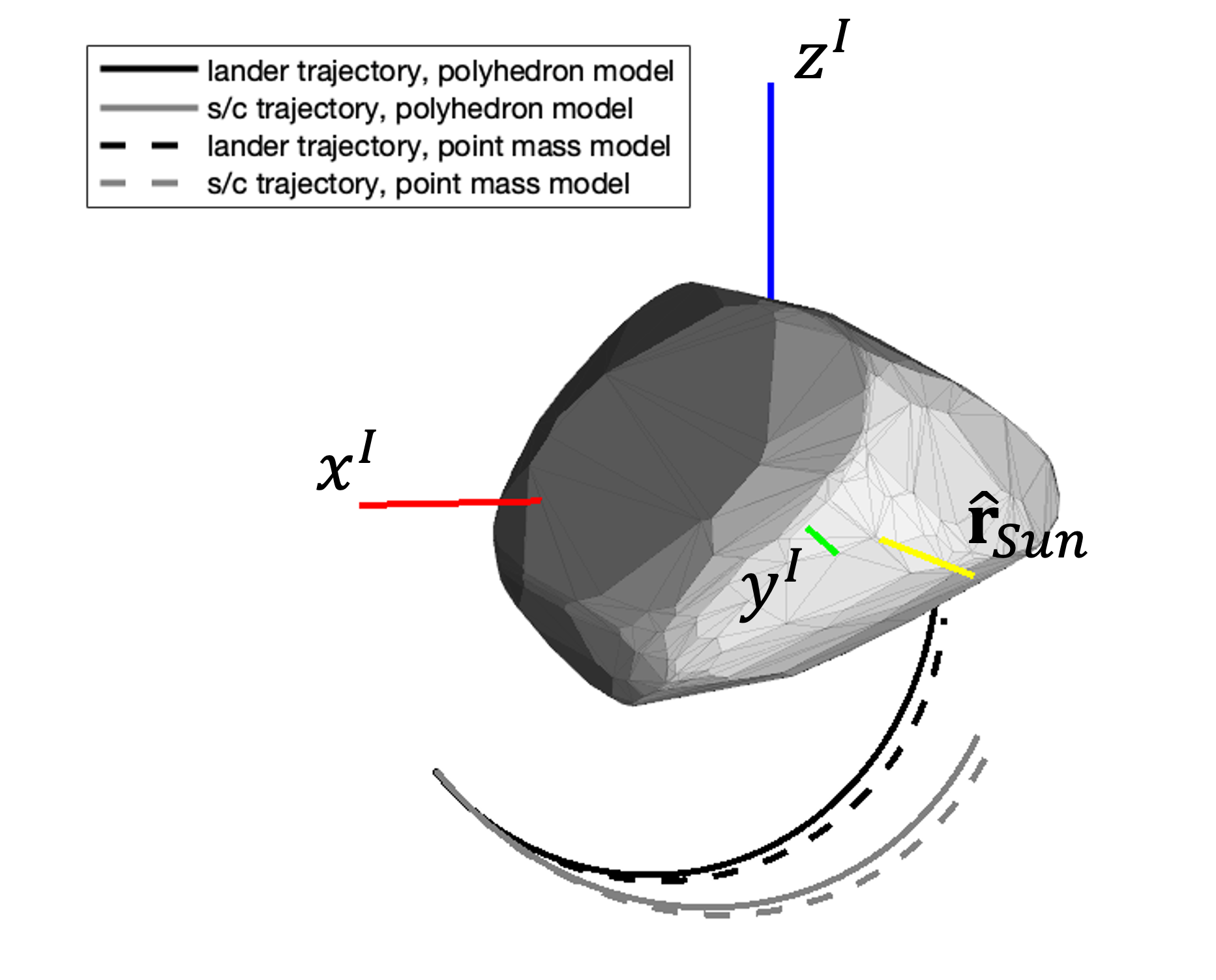}
  \caption{Probe and main S/C post-deployment trajectories}
  \label{fig:lander and sc trajectories}
   \end{center}
\end{figure}

Figure~\ref{fig:lander and sc trajectories} depicts an example of the probe descent trajectory and corresponding trajectory of the main S/C for Mid $\mu$ for one of the asteroid attitudes. Let us note that in Fig.~\ref{fig:lander and sc trajectories} Justitia attitude corresponds to the touchdown epoch. 

\begin{figure}[h!]
 \begin{center}
  \includegraphics[width=1\textwidth]{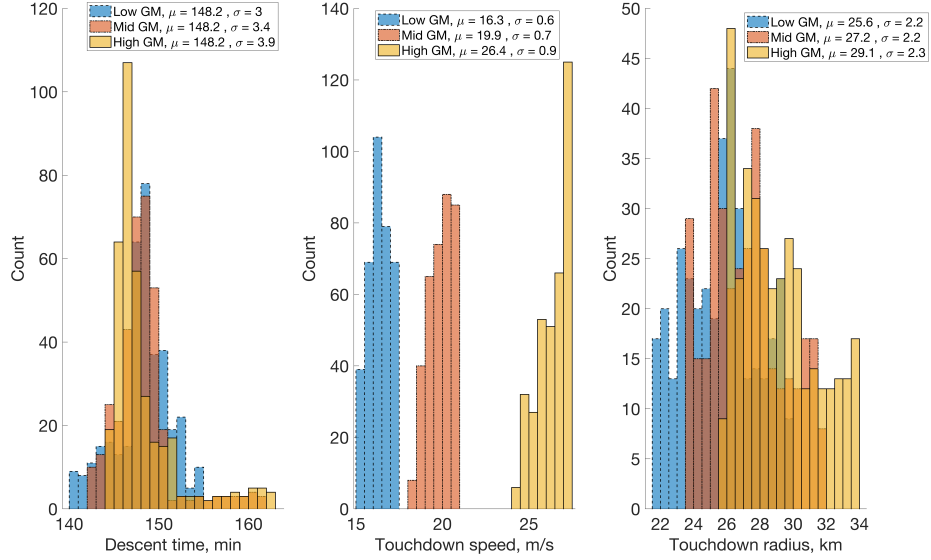}
  \caption{Statistics of descent trajectories calculated using the polyhedron gravity model for different initial asteroid orientations in the Low, Mid, and High $\mu$ cases: descent time, touchdown speed, and touchdown radius}
  \label{fig:HF trajectories statistics}
   \end{center}
\end{figure}

Figure~\ref{fig:HF trajectories statistics} illustrates the statistics of the main parameters of the probe descent trajectory such as the descent time, touchdown speed, and touchdown radius for three scenarios of asteroid density $\rho$ and various initial asteroid attitudes for a range of asteroid rotation angle $\phi$ (see Eq.~\ref{eq:asteroid angles}) from 0 to $2\pi$. The trajectories are propagated using the polyhedron gravity model to take into account the asteroid shape effects, while maintaining the assumption of uniform bulk density.

The results illustrated in Fig.~\ref{fig:HF trajectories statistics} demonstrate that, despite the irregular gravity field and the asteroid rotation, the descent time remains close to the targeted 2.5 hours for all considered Justitia $\mu$ values. The touchdown speed shows a noticeable dependence on the assumed gravitational parameter, increasing for the high-$\mu$ case due to stronger gravitational acceleration along the descent arc. The dispersion in touchdown radius relative to the point-mass model reflects the influence of gravity field irregularity, shape, and asteroid rotation. Nevertheless, the deviations remain bounded and do not prevent surface intersection from occurring within the first revolution, as assumed. The obtained dispersion demonstrates that the selected deployment orbit family remains dynamically feasible despite variations in the assumed gravitational parameter and asteroid rotational phase. While touchdown radius and impact velocity exhibit sensitivity to the asteroid configuration, the descent duration remains close to the targeted value for all considered cases. This indicates that the proposed deployment strategy is sufficiently robust for preliminary mission design and imaging performance assessment.

\section{Asteroid Surface Coverage Analysis}
\label{sec: Surface Coverage Analysis}

The section investigates the asteroid imaging and surface coverage performance achievable during the ballistic descent of the EMA probe. The analysis aims to evaluate the geometrical observability of asteroid (269) Justitia under realistic descent trajectories, illumination conditions, and probe attitude motion regimes. Rather than performing photorealistic rendering or detailed image synthesis, the study focuses on a first-order geometrical assessment of the visible and illuminated asteroid surface regions observable by the onboard payload cameras throughout descent.

Coverage and visibility analysis approaches based on ray tracing, polyhedron intersection, and camera FOV discretization are commonly used in small-body proximity operations and optical navigation studies \cite{SISPO2020, PUGLIATTI202314}. The approach is also conceptually related to the point coverage models previously used for Earth coverage assessment and distributed spacecraft observation problems~\cite{Mortari2004Flower, aerospace9080419}. In the present work, the asteroid surface is represented by a polyhedron model, while the payload camera FOV is discretized using a set of nearly uniformly distributed rays. The visibility of illuminated surface regions is evaluated through ray-facet intersection analysis performed along the simulated descent trajectories.

Particular attention is devoted to the influence of the probe rotational motion and attitude control approach on the achievable imaging performance. Three representative cases are considered: passive uncontrolled rotational motion after deployment, Sun-spin control mode, and asteroid-pointing mode. The corresponding simulations are used to qualitatively and quantitatively assess achievable asteroid observation conditions, surface coverage, and ground sampling distance throughout the ballistic descent phase.

\subsection{Coverage Analysis Approach} 

The surface coverage analysis presented in this section utilizes the probe translational and rotational motion states obtained in Section~\ref{sec:deployment orbit selection} and~\ref{app: attitude control algorithms}. The probe position vector $\mathbf{r}(t)$ and attitude unit quaternion $\mathbf{q}^{\mathcal{IB}}(t)$ are assumed to be known from the simulated descent dynamics and attitude control models. These data are used to evaluate the observable asteroid surface regions, illumination conditions, and imaging performance of the payload cameras throughout descent.

The asteroid is represented by the polyhedron shape model introduced in Section~\ref{sec:Justitia properties}, while its rotational motion is described by the time-varying direction cosine matrix $\mathbf{C}^{\mathcal{IA}}(t)$ defined in Eq.~\ref{eq:asteroid angles}. The payload cameras are characterized by their half FOV angles $\gamma$, angular resolutions $\gamma_{\mathrm{res}}$, and boresight direction vectors defined in the probe body-fixed frame $\mathcal{F}^\mathcal{B}$ (see Section~\ref{sec: Lander description}).

Unlike conventional point coverage models, the present analysis aims to evaluate the geometrically observable illuminated surface regions of the asteroid, particularly for irregular small bodies with complex local topography, rather than only verifying whether the asteroid is located inside the camera FOV. This distinction is important because the camera boresight may be arbitrarily oriented relative to the asteroid, while portions of the observed surface can remain shadowed or only partially visible. Therefore, a geometrical coverage analysis method is developed to estimate the actual observable surface regions corresponding to each camera observation.

The proposed method discretizes the camera FOV using a set of nearly uniformly distributed rays generated inside the camera viewing cone. For each simulation time instant, ray--facet intersection tests are performed against the asteroid polyhedron model. A facet is considered observable only if the corresponding observation condition is satisfied and the ray intersection represents the closest visible surface point along the ray direction. The obtained intersection points are then used to reconstruct an approximate observable surface patch represented by a triangulated surface mesh, allowing partial coverage of individual asteroid facets to be taken into account. Finally, illumination conditions are evaluated for the reconstructed surface region using the Sun direction vector, while the corresponding local ground sampling distance (GSD) values are estimated for the observed surface elements.

\begin{figure}[h!]
 \begin{center}
  \includegraphics[width=0.8\textwidth]{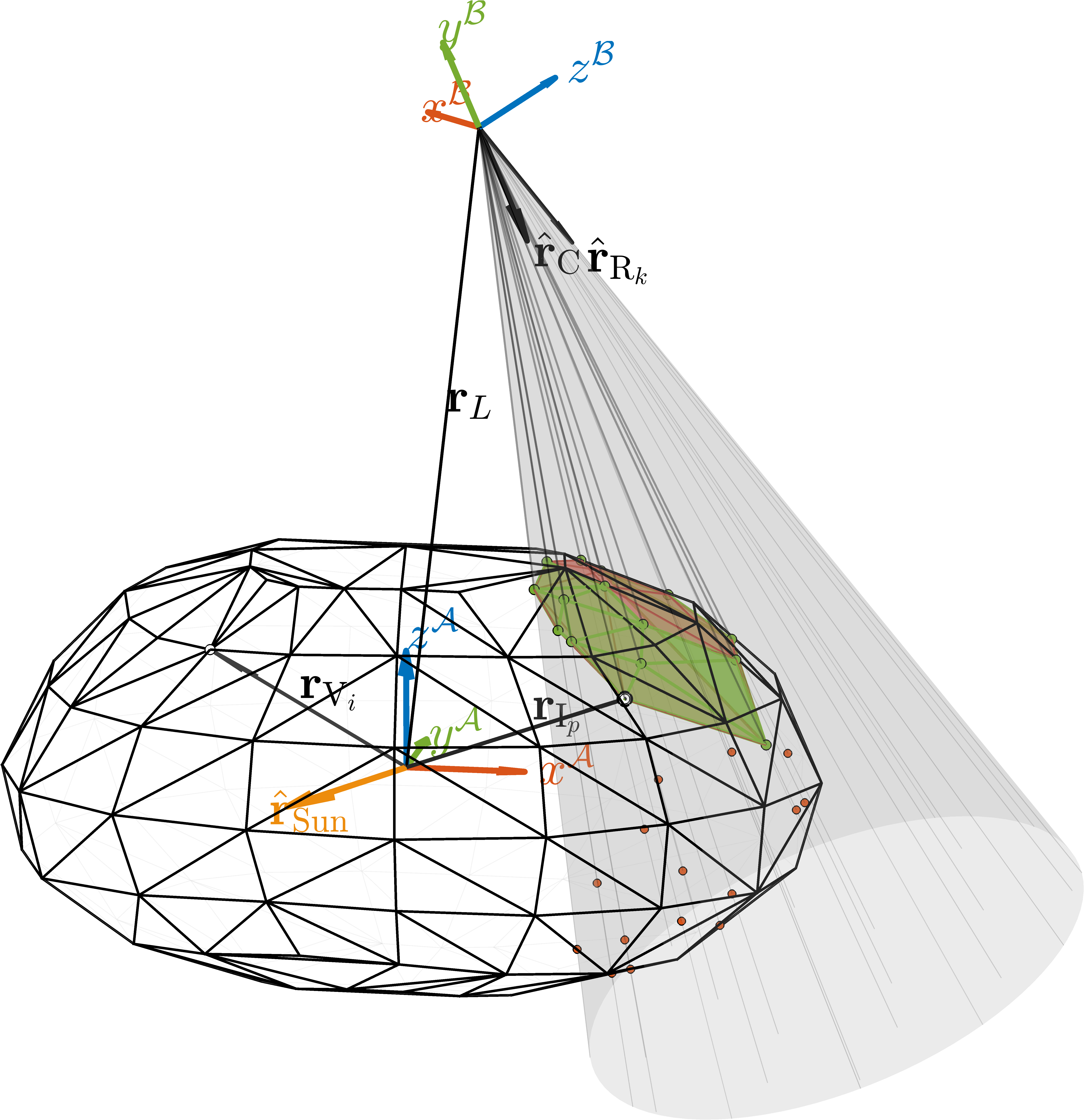}
  \caption{Polyhedron-shape asteroid coverage geometry}
  \label{fig:asteroid coverage geometry}
 \end{center}
\end{figure}

Figure~\ref{fig:asteroid coverage geometry} illustrates the proposed geometrical coverage analysis method. The corresponding workflow is summarized as follows.

\begin{enumerate}

    \item For each payload camera, the FOV is pre-discretized by a set of ray direction vectors $\hat{\mathbf{r}}_{\mathrm{R}_k}$. The rays inside the camera viewing cone are generated using a Fibonacci spiral-based distribution, while additional rays are uniformly distributed on the cone surface.

    \item At each simulation time step, a necessary geometrical condition for the camera to observe the asteroid is first checked:
    
    \begin{equation}\label{eq:necessary observation condition_new}
        \angle \big( \hat{\mathbf{r}}_{\mathrm{c}}, \hat{\mathbf{r}}_{\mathrm{ast}} \big) < \beta_{\mathrm{crit}}^{\ast},
    \end{equation}
    where
    \begin{equation}
        \beta_{\mathrm{crit}}^{\ast} = \arcsin \left( \frac{r_{\mathrm{V,max}}}{r_{\mathrm{L}}} \right) + \gamma,
        \qquad
        r_{\mathrm{V,max}} = \max_i \left\| \mathbf{r}_{\mathrm{V}_i} \right\|.
    \end{equation}
    
    If the condition is not satisfied, the camera is assumed not to observe the asteroid at the given time instant and the ray--facet intersection analysis is skipped. Let us note that the parameter $\beta_{\mathrm{crit}}^{\ast}$ in contrast to the one used in the orbit attitude selection considers the polyhedron shape of the asteroid instead of spherical one.

    \item If the condition is satisfied, the precomputed ray directions are transformed into the common reference frame using the current probe attitude unit quaternion. For each ray direction $\hat{\mathbf{r}}_{\mathrm{R}_k}$ and each asteroid facet, the ray--facet intersection point $\mathbf{r}_{\mathrm{I}_p}$ is calculated. If several valid intersections are found along the same ray direction, the closest intersection point to the probe is selected. In Fig.~\ref{fig:asteroid coverage geometry}, the selected closest intersection points are shown in green, while the remaining non-selected intersection points are shown in red.

    \item The obtained intersection points $\mathbf{r}_{\mathrm{I}_p}$ are connected to reconstruct an approximate observed surface patch represented by a triangular mesh. The normal vectors $\hat{\mathbf{n}}_{\mathrm{F}_j}$ are then calculated for the reconstructed mesh facets.

    \item For each reconstructed surface facet, the illumination condition is evaluated using the facet outward normal vector $\hat{\mathbf n}_{\mathrm{F}_j}$ and the Sun direction vector $\hat{\mathbf r}_{\mathrm{Sun}}$. A facet is considered illuminated if

    \begin{equation}
    \hat{\mathbf n}_{\mathrm{F}_j} \cdot
    \hat{\mathbf r}_{\mathrm{Sun}}
    >0.
    \end{equation}
    
    The areas of all illuminated facets are summed to estimate the illuminated observable surface area.

    \item The average GSD of the asteroid image is estimated as
        \begin{equation}
            \mathrm{GSD} \approx 
            \gamma_{\mathrm{res}}
            \,
            \mathrm{avg}_p
            \left(
            \left\|
            \mathbf{r}_{\mathrm{I}_p}
            -
            \mathbf{r}_{\mathrm{L}}
            \right\|
            \right).
        \end{equation}
\end{enumerate}

All vectors participating in the geometrical coverage analysis should be represented in a common reference frame. For compactness of notation, the explicit frame superscripts are omitted within the coverage analysis procedure whenever the corresponding coordinate representation is clear from the context. The ray--facet intersection analysis utilizes standard ray--triangle intersection procedures from computational geometry~\cite{Ericson2005}. The observable surface patch reconstruction from the obtained intersection points is performed using standard surface triangulation routines available in MATLAB. The developed methodology is subsequently applied to analyze imaging performance for different probe attitude motion regimes.

\subsection{Asteroid Surface Coverage Results}

The developed coverage analysis methodology is applied to three representative attitude motion scenarios: uncontrolled tumbling, Sun-spin stabilization, and asteroid-pointing mode. The objective of the analysis is to evaluate the achievable asteroid observation opportunities during ballistic descent and to assess the influence of the attitude motion strategy on imaging performance.

For the uncontrolled tumbling case, Monte Carlo simulations are performed using a randomly generated initial angular velocity vector with components uniformly distributed within the interval $[-2,2]$ deg/s for each body-fixed axis. For the Sun-spin and asteroid-pointing modes, the probe is assumed to start the descent in the corresponding nominal attitude state. The transient attitude acquisition phase is not considered, as the objective of the analysis is to evaluate the achievable imaging performance of the operational modes rather than the convergence properties of the attitude control algorithms.

The Mid $\mu$ scenario is selected as the reference case for the detailed analysis presented below. The Low and High $\mu$ surface coverage cases were also analyzed; however, they exhibited qualitatively similar trends and did not provide additional insight into the effect of the attitude strategy. Therefore, only the Mid $\mu$ case is discussed in detail. Consequently, the imaging performance is found to be governed primarily by the selected attitude strategy rather than by the considered variations of the asteroid gravitational parameter.

To quantify asteroid observation performance, several complementary metrics are introduced.

The total observation availability is defined as

\begin{equation}
P_{\mathrm{obs}}
= \frac{T_{\mathrm{obs}}}
{T_{\mathrm{desc}}},
\end{equation}

\noindent where $T_{\mathrm{obs}}$ is the cumulative duration during which a non-zero asteroid surface area is visible within the payload camera field of view, and $T_{\mathrm{desc}}$ is the total descent duration.

Similarly, the illuminated observation availability is defined as

\begin{equation}
P_{\mathrm{obs,lit}}
= \frac{T_{\mathrm{obs,lit}}}
{T_{\mathrm{desc}}},
\end{equation}

where $T_{\mathrm{obs,lit}}$ is the cumulative duration during which a non-zero illuminated asteroid surface area is visible by the payload camera.

The observation interval, denoted by $\tau_{\mathrm{obs}}$, is defined as a continuous time interval during which the asteroid surface remains visible within the camera field of view. The illuminated observation interval is defined analogously by considering only the illuminated visible surface. The revisit interval, denoted by $\tau_{\mathrm{rev}}$, is defined as a continuous time interval between two subsequent observation intervals during which no asteroid surface is visible within the camera field of view. In addition, the instantaneous illuminated observed area and average image ground sampling distance (GSD) are evaluated throughout the descent.

The selected metrics are adapted from conventional ground coverage analysis and modified for the present asteroid descent problem. Since the considered attitude regimes produce different observation patterns, the performance is not characterized by a single metric. Observation availability, observation intervals, and revisit intervals are used to describe the temporal structure and periodicity of asteroid visibility, while the illuminated observed area and GSD characterize the geometrical properties of the observed surface region for representative cases. This combination of metrics allows the comparison of stochastic tumbling, periodic Sun-spin observations, and continuous asteroid-pointing observations without overloading the analysis with redundant data.

\subsubsection{Uncontrolled Tumbling}

\begin{figure}[!ht]
\centering
\includegraphics[width=\linewidth]{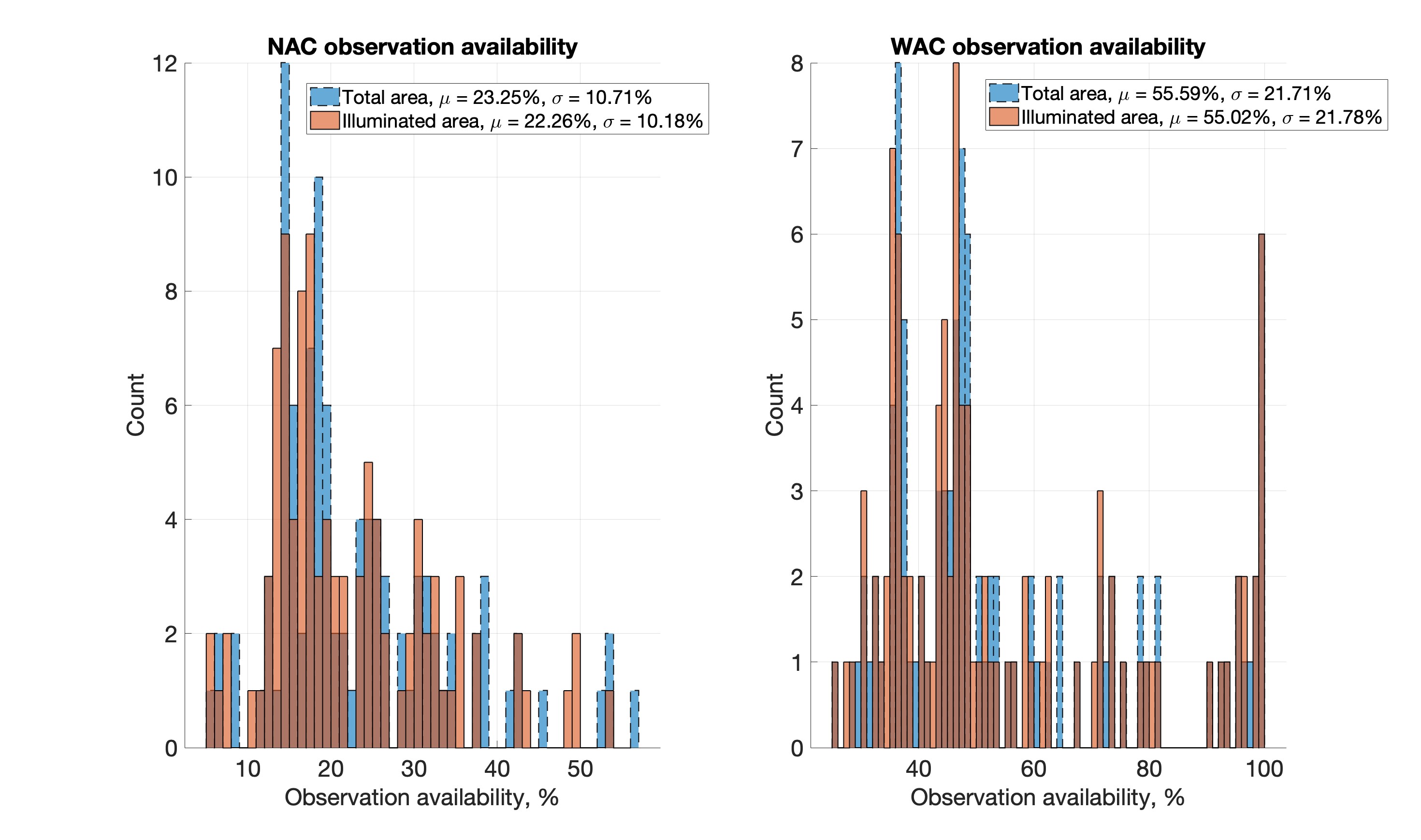}
\caption{Observation availability statistics for the uncontrolled tumbling case (Mid $\mu$ scenario)}
\label{fig:tumbling_availability}
\end{figure}

The observation availability statistics for the uncontrolled tumbling case are presented in Fig.~\ref{fig:tumbling_availability}. The results demonstrate that useful asteroid observations remain possible even without active attitude control. However, the observation process strongly depends on the initial rotational state of the probe, which leads to significant variability between Monte Carlo cases.

\begin{figure}[!ht]
\centering
\includegraphics[width=\linewidth]{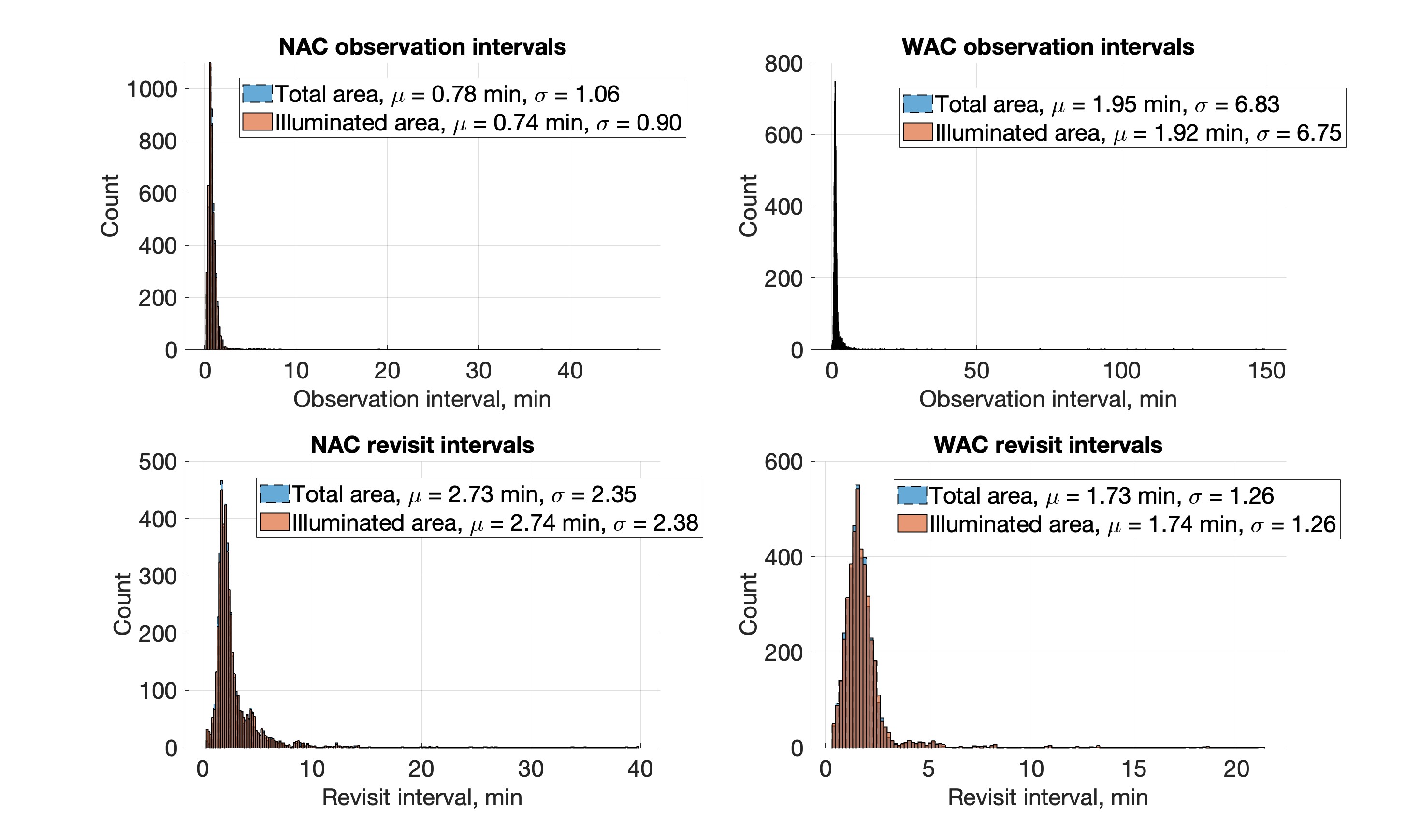}
\caption{Observation and revisit interval statistics for the uncontrolled tumbling case (Mid $\mu$ scenario)}
\label{fig:tumbling_intervals}
\end{figure}

The temporal characteristics of the observation process are further illustrated in Fig.~\ref{fig:tumbling_intervals}. The results show that passive tumbling may produce both useful observation intervals and long gaps when the asteroid remains outside the camera field of view. Therefore, although some cases provide favorable observation opportunities throughout descent, passive tumbling does not guarantee stable asteroid visibility.

\begin{figure}[!ht]
\centering
\includegraphics[width=\linewidth]{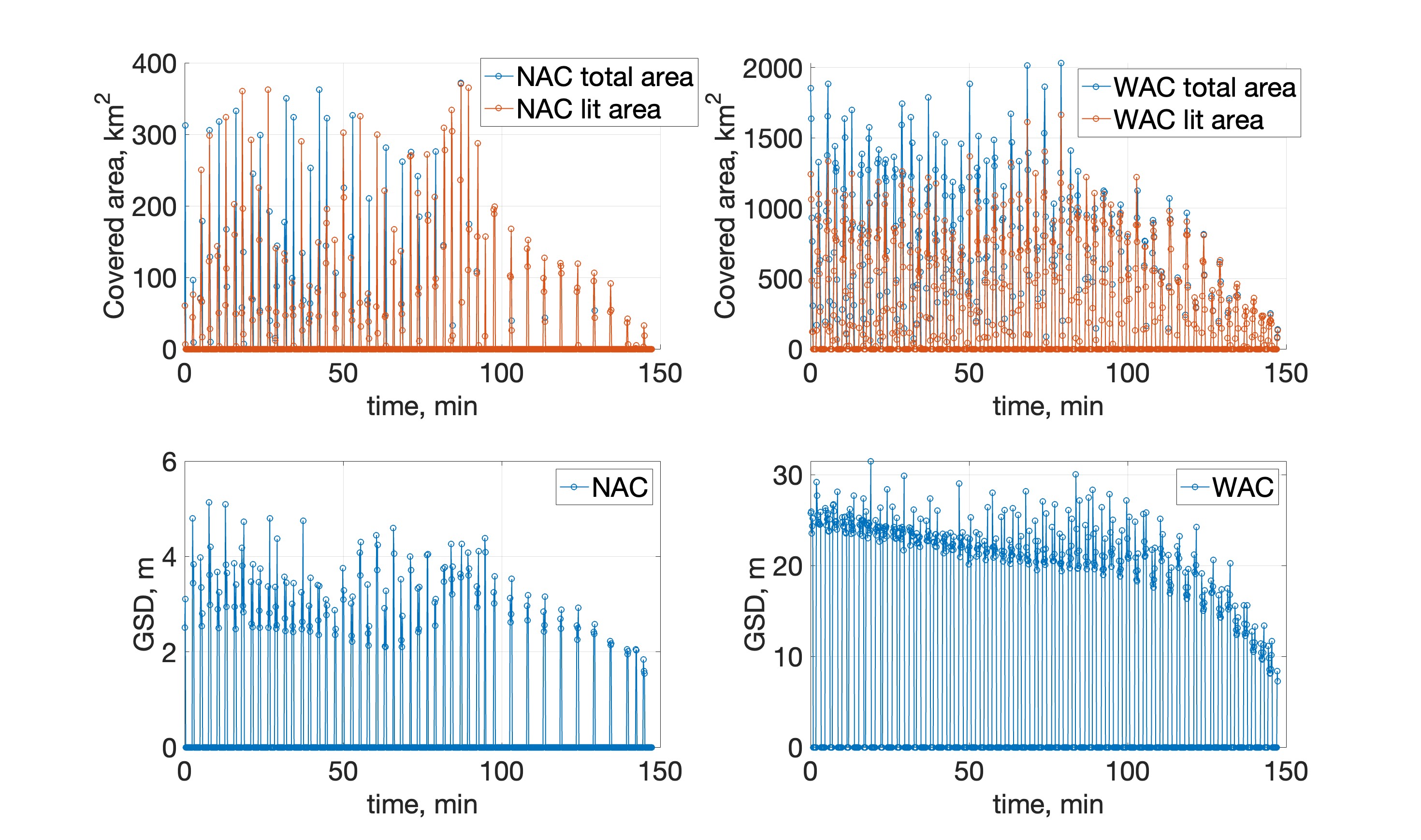}
\caption{Representative asteroid coverage and GSD evolution during ballistic descent for the uncontrolled tumbling case}
\label{fig:tumbling_example}
\end{figure}

A representative descent case is presented in Fig.~\ref{fig:tumbling_example}. Although the image resolution improves as the probe approaches the asteroid surface, the uncontrolled rotational motion produces discontinuous observation opportunities. These results indicate that passive tumbling should be regarded primarily as a contingency operating condition rather than a nominal science observation mode.

\subsubsection{Sun-Spin Mode}

The Sun-spin mode is one of the active attitude control regimes and serves as a safe operational mode for the probe ADCS. In this mode, the probe maintains Sun pointing and performs a slow rotation around the Sun direction, which results in periodic asteroid observation opportunities. Three representative spin rates of 0.5, 1, and 2 deg/s are investigated.

\begin{figure}[!ht]
\centering
\includegraphics[width=\linewidth]{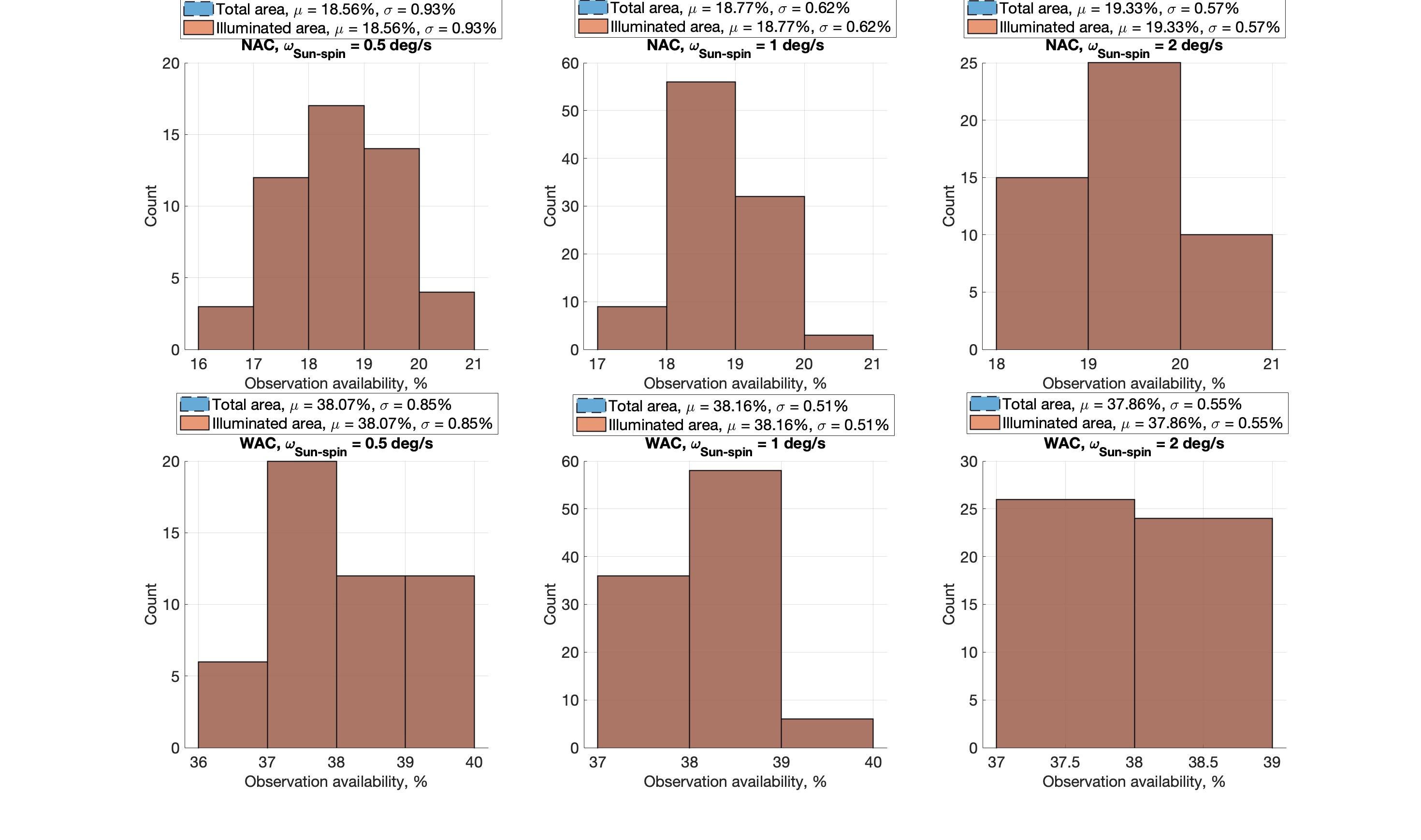}
\caption{Observation availability statistics for Sun-spin mode with different spin rates}
\label{fig:sunspin_availability}
\end{figure}

The corresponding observation availability statistics are presented in Fig.~\ref{fig:sunspin_availability}. In contrast to the uncontrolled tumbling case, the observation process becomes deterministic and repeatable. The overall observation availability remains of the same order as in the tumbling scenario; however, the asteroid is revisited periodically throughout the descent as a direct consequence of the controlled rotational motion.

\begin{figure}[!ht]
\centering
\includegraphics[width=\linewidth]{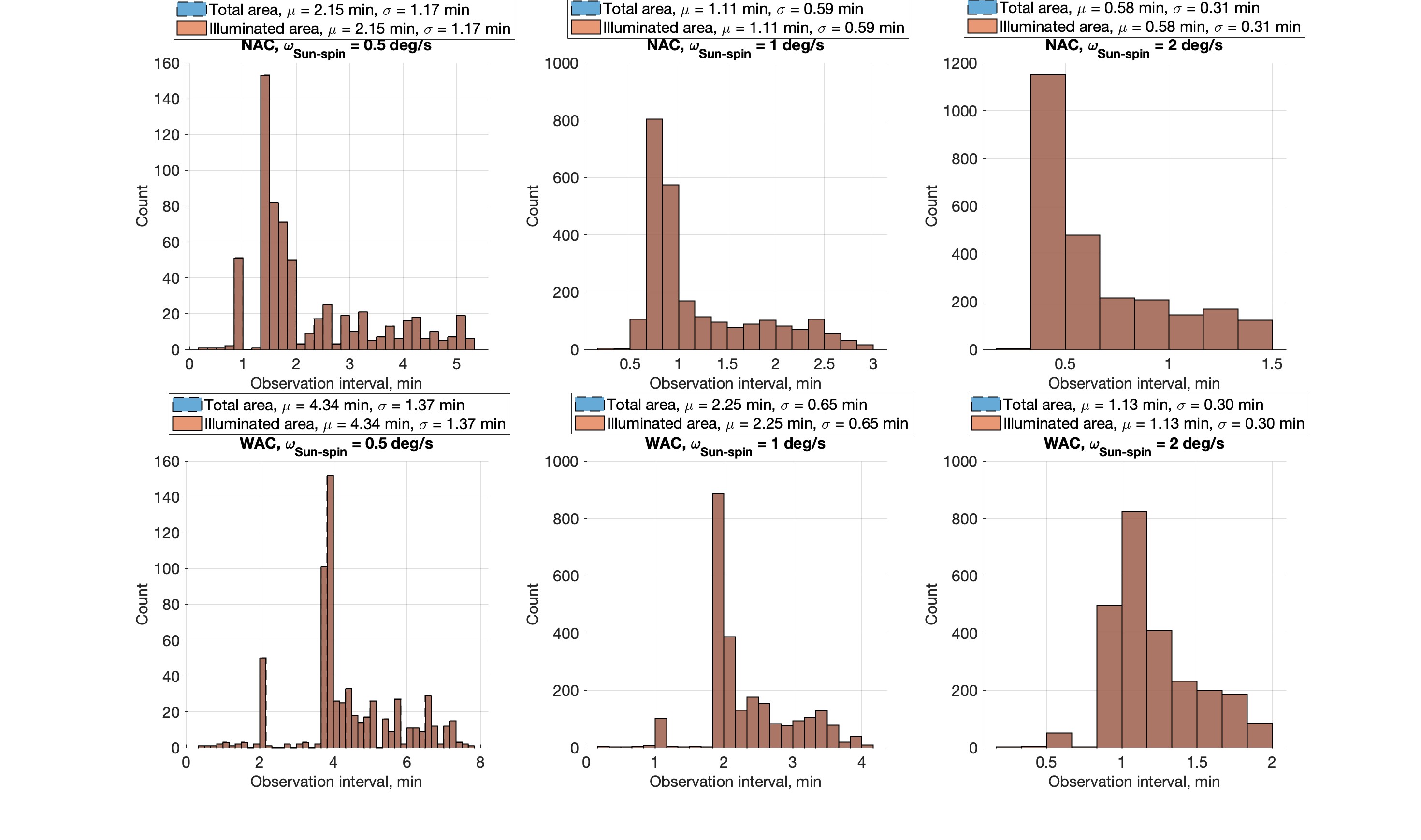}
\caption{Observation interval for Sun-spin mode}
\label{fig:sunspin_observation_intervals}
\end{figure}

The temporal characteristics of the observations are illustrated in Fig.~\ref{fig:sunspin_observation_intervals}. Compared with uncontrolled tumbling, the obtained distributions become considerably narrower, reflecting the periodic nature of the Sun-spin regime. The spin rate directly influences the balance between observation duration and revisit frequency. Lower spin rates provide longer uninterrupted observation intervals but are accompanied by longer revisit times, whereas higher spin rates generate shorter observation windows while increasing the frequency with which the asteroid re-enters the camera field of view.

\begin{figure}[!ht]
\centering
\includegraphics[width=\linewidth]{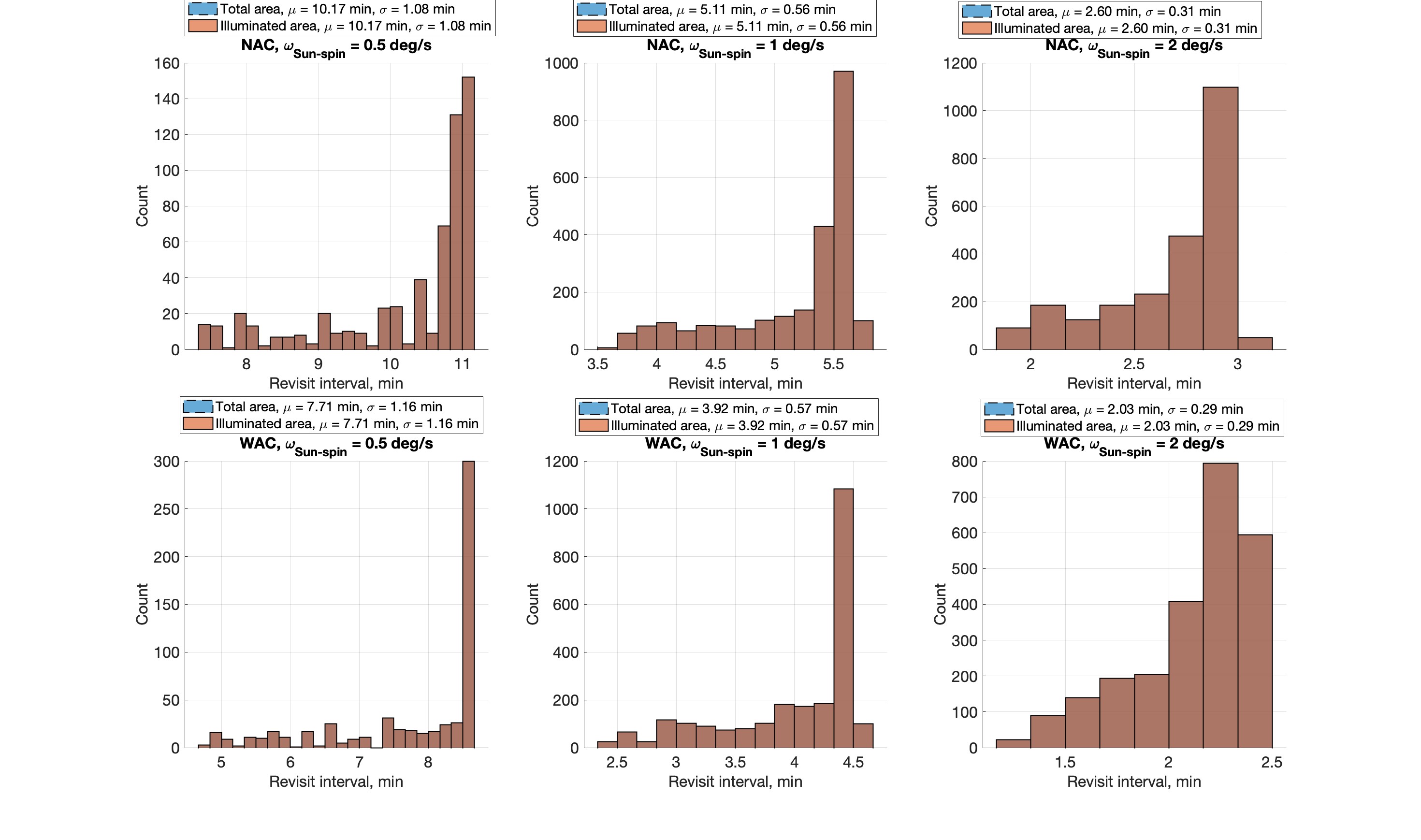}
\caption{Revisit intervals for Sun-spin mode}
\label{fig:sunspin_revisit_intervals}
\end{figure}

The revisit interval statistics shown in Fig.~\ref{fig:sunspin_revisit_intervals} further demonstrate the deterministic behavior of the Sun-spin mode. Unlike the broad distributions observed in the uncontrolled tumbling case, the revisit intervals remain concentrated around characteristic values determined primarily by the selected spin rate.

\begin{figure}[!ht]
\centering
\includegraphics[width=\linewidth]{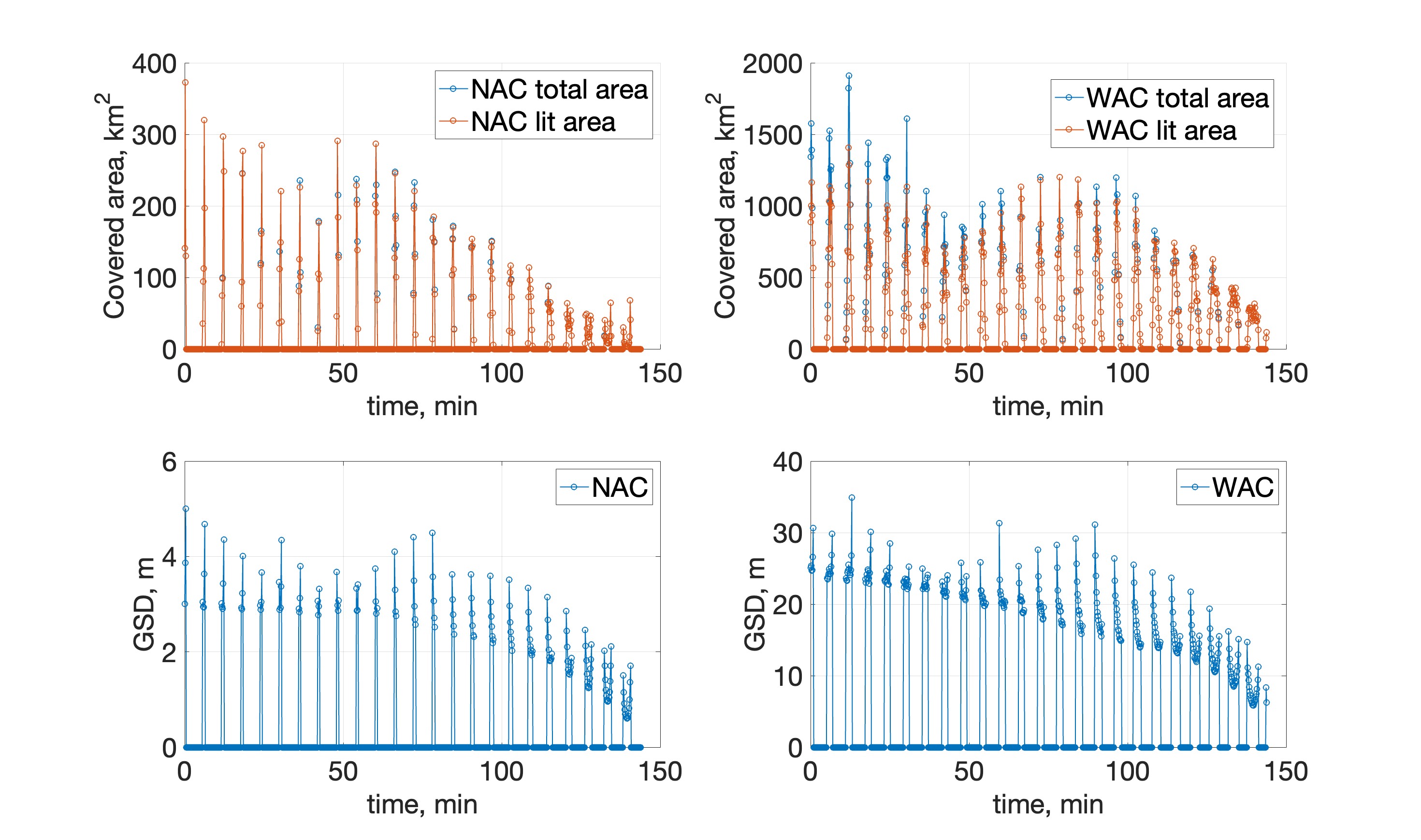}
\caption{Representative asteroid coverage and GSD evolution during ballistic descent for the 1 deg/s Sun-spin mode}
\label{fig:sunspin_example}
\end{figure}

A representative descent case for the reference 1 deg/s spin rate is presented in Fig.~\ref{fig:sunspin_example}. The periodic observation pattern is clearly visible throughout the descent. Consequently, the principal advantage of Sun-spin stabilization is not a substantial increase in observation availability, but rather the transformation of the observation process from a stochastic regime into a predictable periodic regime. This property makes Sun-spin mode particularly attractive as a safe operational mode capable of providing regular imaging opportunities while maintaining continuous Sun-referenced attitude knowledge.

\subsubsection{Asteroid-Pointing Mode}

The asteroid-pointing mode represents the nominal science observation scenario. In this mode, the probe maintains Sun pointing while aligning the payload camera direction with the asteroid direction projected onto the plane orthogonal to the Sun direction, according to the target attitude defined in Eq.~\ref{eq:asteroid pointing attitude}.

The obtained results demonstrate that asteroid visibility is maintained for nearly the entire descent duration. In contrast to the previous scenarios, the observation process becomes largely independent of the initial rotational state and is only weakly affected by variations in asteroid rotational phase. Consequently, asteroid-pointing mode provides both the highest observation availability and the most reliable imaging conditions among all investigated attitude control modes.

To further characterize the achievable imaging performance, Fig.~\ref{fig:asteroid pointing, area and gsd statistics} presents the mean and standard deviation of the total and illuminated observed area and image GSD obtained from Monte Carlo simulations. The results indicate not only improved average performance but also reduced variability compared with passive tumbling. The observed area tends to decrease during descent because the fixed camera field of view covers a smaller physical surface region as the probe approaches the asteroid. At the same time, the average GSD improves as the observation distance decreases, reflecting the expected trade-off between surface coverage area and spatial resolution. The relatively small dispersion of both observed area and GSD demonstrates that the asteroid-pointing strategy provides robust and repeatable imaging performance despite uncertainties associated with touchdown location and asteroid rotational phase. Therefore, asteroid-pointing mode is identified as the preferred operational mode for maximizing scientific return during the ballistic descent phase.

\begin{figure}[!ht]
\centering
\includegraphics[width=\linewidth]{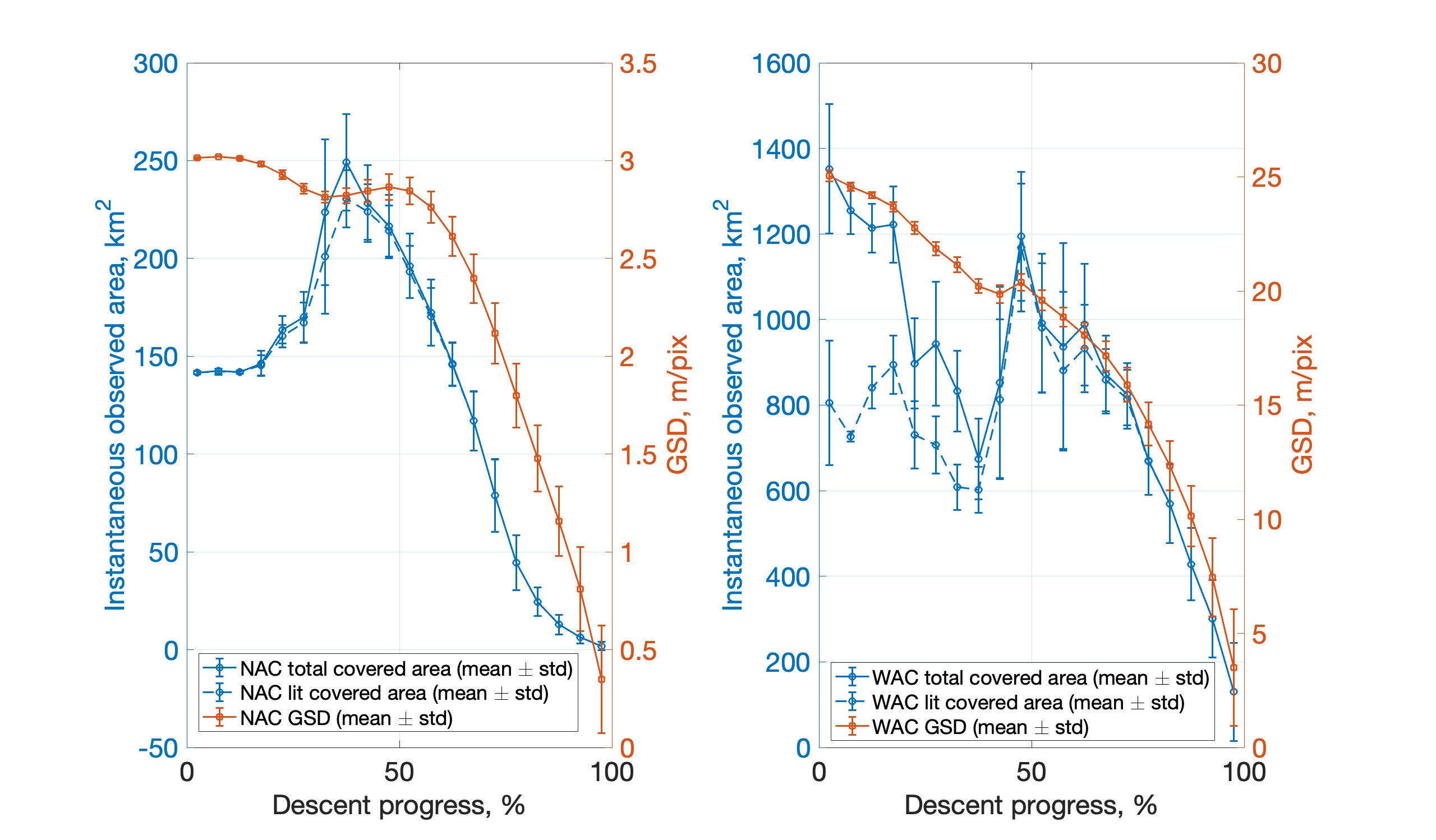}
\caption{Mean and standard deviation of the total and illuminated observed area and image GSD for the asteroid-pointing mode, Mid $\mu$ case}
\label{fig:asteroid pointing, area and gsd statistics}
\end{figure}

\subsubsection{Comparison of Attitude Modes}

Table~\ref{tab:coverage_summary} summarizes the observation performance metrics for all considered attitude modes. For the asteroid-pointing mode, the revisit interval is zero because asteroid visibility is maintained throughout the descent.

\begin{table}[t]
\centering
\caption{Summary of the asteroid surface coverage results for all considered attitude modes. The values are given as mean $\pm$ standard deviation}
\label{tab:coverage_summary}
\footnotesize
\resizebox{\textwidth}{!}{%
\begin{tabular}{p{3.4cm}lcccc}
\toprule
Mode & Camera & $P_{\mathrm{obs}}$ [\%] & $P_{\mathrm{obs,lit}}$ [\%] & $\tau_{\mathrm{obs}}$ [min] & $\tau_{\mathrm{rev}}$ [min] \\
\midrule
\multirow{2}{3.4cm}{Uncontrolled tumbling}
& NAC & $23.25 \pm 10.71$ & $22.26 \pm 10.18$ & $0.78 \pm 1.06$ & $2.73 \pm 2.35$ \\
& WAC & $55.59 \pm 21.71$ & $55.02 \pm 21.78$ & $1.95 \pm 6.83$ & $1.73 \pm 1.26$ \\

\multirow{2}{3.4cm}{Sun-spin, 0.5 deg s$^{-1}$}
& NAC & $18.56 \pm 0.93$ & $18.56 \pm 0.93$ & $2.15 \pm 1.17$ & $10.17 \pm 1.08$ \\
& WAC & $38.07 \pm 0.85$ & $38.07 \pm 0.85$ & $4.34 \pm 1.37$ & $7.71 \pm 1.16$ \\

\multirow{2}{3.4cm}{Sun-spin, 1 deg s$^{-1}$}
& NAC & $18.77 \pm 0.62$ & $18.77 \pm 0.62$ & $1.11 \pm 0.59$ & $5.11 \pm 0.56$ \\
& WAC & $38.16 \pm 0.51$ & $38.16 \pm 0.51$ & $2.25 \pm 0.65$ & $3.92 \pm 0.57$ \\

\multirow{2}{3.4cm}{Sun-spin, 2 deg s$^{-1}$}
& NAC & $19.33 \pm 0.57$ & $19.33 \pm 0.57 $ & $0.58 \pm 0.31$ & $2.60 \pm 0.31$ \\
& WAC & $37.86 \pm 0.55 $ & $37.86 \pm 0.55$ & $1.13 \pm 0.30$ & $2.03 \pm 0.29$ \\

\multirow{2}{3.4cm}{Asteroid-pointing}
& NAC & $100 $ & $100 $ & $148.2 \pm 3.4$ & 0 \\
& WAC & $100 $ & $100 $ & $148.2 \pm 3.4$ & 0 \\
\bottomrule
\end{tabular}%
}
\end{table}

The results reveal a clear progression in observation quality as the level of attitude control increases. Passive tumbling produces a stochastic observation process characterized by large variability in both observation availability and observation continuity. Although useful observations remain possible, imaging opportunities strongly depend on the initial rotational state and cannot be guaranteed during critical phases of the descent.

The Sun-spin mode transforms the observation process into a deterministic periodic regime. While the overall observation availability remains comparable to the tumbling case, observation opportunities become predictable and repeatable. Furthermore, the spin rate provides a direct means of adjusting the trade-off between observation interval duration and revisit frequency. Lower spin rates maximize uninterrupted observation periods, whereas higher spin rates increase the temporal sampling frequency of the asteroid surface.

The asteroid-pointing mode provides the best overall performance. Continuous payload alignment with the asteroid results in nearly uninterrupted observations throughout descent, maximized surface visibility, and the most consistent imaging conditions. In addition to achieving the highest observation availability, this mode also minimizes variability in observed area and image resolution.

Overall, the analysis demonstrates that the selected attitude strategy has a substantially greater influence on imaging performance than the considered uncertainty in the asteroid gravitational parameter. The obtained results therefore identify asteroid-pointing as the preferred operational mode for maximizing scientific return during the EMA Justitia probe descent phase, while Sun-spin mode represents a practical backup strategy capable of providing predictable periodic observations.

\section{Conclusion}

The present work develops a consistent mission design framework for the EMA Justitia probe, whose science observations are to be performed during an uncontrolled ballistic descent. The probe communication architecture, uncertainty in the asteroid physical parameters, and the operational and safety constraints of the main spacecraft mean that the probe cannot rely on prior knowledge of its exact descent trajectory at the time of deployment. These limitations were therefore transformed into formal mission-level requirements on deployment orbit geometry, illumination, attitude determination and control, and payload pointing conditions.

A deployment orbit selection method was developed in the terminator reference frame. The method uses the minimum boresight deviation margin, $\Delta\beta_{\min}$, to identify families of deployment orbit orientations that maintain the asteroid within the payload camera FOV while providing suitable illumination conditions. This formulation enables the final deployment orbit to be selected during the main spacecraft proximity operations phase without requiring the probe to have prior knowledge of the selected orbit. The selected near-terminator geometry also provides continuous Sun visibility and supports the proposed attitude control approach.

Reference deployment orbits were selected for the considered Low, Mid, and High values of the Justitia gravitational parameter. Their geometry was verified against the derived illumination and observation requirements, and the corresponding descent trajectories were propagated using a polyhedron gravity model. The results show that the selected orbit family remains feasible for the considered gravity cases and asteroid rotational phases, while maintaining descent durations close to the prescribed value of 2.5 h. Variations in the gravitational parameter and asteroid attitude mainly affect the touchdown location and impact conditions rather than the overall descent duration and observation geometry.

The surface coverage analysis further demonstrated the strong coupling between the deployment trajectory and the probe attitude motion. Uncontrolled tumbling can provide useful but stochastic observation opportunities. The Sun-spin mode converts this behavior into a predictable periodic observation regime and therefore represents a viable safe mode. The three-axis asteroid-pointing mode provides the most stable imaging conditions, the most repeatable illuminated surface coverage, and the most favorable GSD evolution, and is consequently identified as the preferred science observation mode. Together, these results confirm that the selected deployment orbit family and attitude control concept provide suitable conditions for ballistic descent-based asteroid imaging.

The present analysis is based on the current mission design maturity and therefore includes several simplifying assumptions. The asteroid is represented by a uniform-density polyhedron model, the deployment-state and navigation uncertainties are not yet propagated through the analysis, and the attitude control assessment assumes idealized Sun and asteroid direction measurements. The surface coverage analysis also uses simplified asteroid-referenced pointing definitions and does not yet include detailed sensor and actuator models. These aspects will be addressed in subsequent studies.

Overall, the study demonstrates that the initial probe and main spacecraft constraints can be translated into a formal deployment orbit selection problem and addressed through a coupled analysis of orbit geometry, attitude control, and surface coverage. The proposed approach provides a mission design basis for the EMA Justitia probe and a general methodology for future ballistic asteroid probes operating with limited onboard knowledge of their final descent trajectory.

At the time of this writing (Fall 2026), the Probe was undergoing flight qualification testing, including environmental and performance tests. The Probe is planned for delivery, followed by integration with the MBR Explorer and subsequent environmental testing of the main S/C. The mission is scheduled for launch in 2028, with arrival at (269) Justitia planned for 2034.

\section*{Acknowledgment}

The development of the probe for the Emirates Mission to the Asteroid Belt (EMA) is supported by the United Arab Emirates Space Agency.

\appendix
\section{Polyhedron Gravity Model}
\label{app: polyhedron gravity model}

The polyhedron model of the gravity field from~\cite{werner1996exterior} is described as follows:

\begin{multline} \label{eq:gravity field}
    U(\mathbf{r}) = \frac{1}{2}G\rho \sum_{e \in edges} \textbf{r}_e^{\top} \cdot \textbf{E}_e \cdot \textbf{r}_e  L_e -
    \frac{1}{2}G\rho \sum_{f \in faces} \textbf{r}_f^{\top} \cdot \textbf{F}_f \cdot \textbf{r}_f ~ \omega_f,
\end{multline}

\noindent where $\mathbf{r}_e$ is a vector from  a satellite position $\mathbf{r}$ in $\mathcal{F^A}$ to a vertex connecting edge $e$, $\mathbf{r}_f$ is a vector from satellite position $\mathcal{F^A}$ to a vertex in facet $f$, edges dyads $\mathbf{E}_e$ and facets dyads $\mathbf{F}_f$ are polyhedron properties that can be pre-computed in advance as they are invariant to a constant asteroid shape, per-edge factors $L_e$ and per-facet factors $\omega_f$ requires computation for a given satellite position in $\mathcal{F^A}$ frame.

The attraction $\nabla U$ is calculated as follows:

\begin{multline} \label{eq:attraction}
    \nabla U(\mathbf{r}) = -G\rho \sum_{e \in edges}  \textbf{E}_e \cdot \textbf{r}_e ~ L_e  +
    G\rho \sum_{f \in faces} \textbf{F}_f \cdot \textbf{r}_f ~ \omega_f 
\end{multline}

In order to determine the touchdown condition, the Laplacian of the potential $\nabla^2U$ is considered:

\begin{equation}\label{eq:laplacian}
    \nabla^2 U(\mathbf{r}) = -G\rho \sum_{f \in faces}\omega_f.
\end{equation}

The Laplacian is equal to zero if probe or a field point is outside polyhedron, while equal to $-4\pi G \rho$ when it is inside.

Let us note that the polyhedron model (Eqs.~\ref{eq:gravity field}--\ref{eq:laplacian}) considers probe position vector represented in asteroid body-fixed frame $\mathcal{F^A}$, while the gravity attraction vector $\nabla U(\mathbf{r}^{\mathcal{A}})$ should be converted into the inertial frame $\mathcal{F^I}$ using the DCM matrix $\mathbf{C}^{\mathcal{IA}}(t)$ (see Eq.~\ref{eq:coordinate transformation}) for orbital motion dynamics equation (see Eq.~\ref{eq:equation of motion}).

\section{Preliminary Implementation of Attitude Control Algorithms}
\label{app: attitude control algorithms}

The control algorithms of the EMA probe include two main control modes, Sun-spin and asteroid-pointing. They utilize feedback control laws derived using Lyapunov function method. The particular implementation of the control laws is as follows.

\paragraph{Sun-spin mode} The Sun-spin control mode is a single axis control law~\cite{mashtakov2021single} derived using Lyapunov direct method theorem. The Lyapunov function and corresponding control law are as follows:

\begin{equation}
V_{\mathrm{ss}} = \frac{1}{2}(\boldsymbol{\omega}_{\mathrm{rel}}^{\mathcal{B}}, \mathbf{J}\boldsymbol{\omega}_{\mathrm{rel}}^\mathcal{B}) + k_{\mathrm{ss}}(1 - (\mathbf{n}^\mathcal{B}, \mathbf{C}^{\mathcal{BI}}\mathbf{n}_{\mathrm{ref}}^\mathcal{I})), \ \ k_{\mathrm{ss}} > 0,
\end{equation}

\begin{multline}
    \mathbf{u}_{\mathrm{ctrl}}^\mathcal{B} = -\mathbf{M}_{\mathrm{ext}}^\mathcal{B} + \boldsymbol{\omega}^\mathcal{B} \times \mathbf{J} \boldsymbol{\omega}^\mathcal{B} - \mathbf{J}(\boldsymbol{\omega}_{\mathrm{rel}}^\mathcal{B}\times \mathbf{C}^{\mathcal{BI}}\boldsymbol{\omega}_{\mathrm{ref}}^\mathcal{I}) + \\ \mathbf{J}\mathbf{C}^{\mathcal{BI}}\dot{\boldsymbol{\omega}}_{\mathrm{ref}}^\mathcal{I} - k_{\omega}\boldsymbol{\omega}_{\mathrm{rel}}^\mathcal{B} - k_{\mathrm{ss}} (\mathbf{C}^{\mathcal{BI}}\mathbf{n}_{\mathrm{\mathrm{ref}}}^\mathcal{I}) \times \mathbf{n}^\mathcal{B},
\end{multline}

\noindent where $\boldsymbol{\omega}_{\mathrm{rel}}^{\mathcal{B}} = \boldsymbol{\omega}^{\mathcal{B}} - \mathbf{C}^{{\mathcal{BI}}}\boldsymbol{\omega}_{\mathrm{ref}}^{\mathcal{I}}$, $\mathbf{n}^{\mathcal{B}}$ is a body-fixed unit vector, $\mathbf{n}_{\mathrm{ref}}^{\mathcal{I}}$ is a target inertial direction, $k_{\omega}$, $k_{\mathrm{ss}}$ are control gains.

\paragraph{Asteroid-pointing mode} The asteroid-pointing is a three axis control law for the target attitude given with a DCM $\mathbf{C}^{\mathcal{IB}_{\mathrm{req}}}$ (see Eq.~\ref{eq:asteroid pointing attitude}) and attitude error $\mathbf{C}_{\mathrm{err}} = \mathbf{C}^{\mathcal{BB}_{\mathrm{req}}} = \mathbf{C}^{\mathcal{BI}} \mathbf{C}^{\mathcal{IB}_{\mathrm{req}}}$:

\begin{equation}
\label{eq:Lyapunov function, asteroid pointing}
V_{\mathrm{DCM}} = \frac{1}{2}(\boldsymbol{\omega}_{\mathrm{rel}}^{\mathcal{B}}, \mathbf{J}\boldsymbol{\omega}_{\mathrm{rel}}^{\mathcal{B}}) + k_{\mathrm{a}}(3 - Tr(\mathbf{C}_{\mathrm{err}})), \ \ k_{\mathrm{a}} > 0,
\end{equation}

\begin{multline}
\mathbf{u}_{\mathrm{ctrl}}^{\mathcal{B}} = -\mathbf{M}_{\mathrm{ext}}^{\mathcal{B}} + \boldsymbol{\omega}^{\mathcal{B}} \times \mathbf{J} \boldsymbol{\omega}^{\mathcal{B}} - \mathbf{J}(\boldsymbol{\omega}_{\mathrm{rel}}^{\mathcal{B}}\times \mathbf{C}^{\mathcal{BI}}\boldsymbol{\omega}_{\mathrm{ref}}^{\mathcal{I}}) \\ + \mathbf{J}\mathbf{C}^{\mathcal{BI}}\dot{\boldsymbol{\omega}}_{\mathrm{ref}}^{\mathcal{I}} - k_{\omega}\boldsymbol{\omega}_{\mathrm{rel}}^{\mathcal{B}} - k_{\mathrm{a}} \mathbf{S},    
\end{multline}

\noindent where $\boldsymbol{\omega}_{\mathrm{rel}}^{\mathcal{B}} = \boldsymbol{\omega}^{\mathcal{B}} - \mathbf{C}^{{\mathcal{BI}}}\boldsymbol{\omega}_{\mathrm{ref}}^{\mathcal{I}}$, $\mathbf{S} = \begin{bmatrix} 
(C_{err})_{23} - (C_{err})_{32} \\ 
(C_{err})_{31} - (C_{err})_{13} \\ 
(C_{err})_{12} - (C_{err})_{21} \end{bmatrix}$, $k_{\omega}$, $k_{\mathrm{a}}$ are control gains. The attitude error matrix $\mathbf{C}_{\mathrm{err}}$ is defined as the rotation mapping the required body frame to the current body frame.

\paragraph{Reaction wheels torque allocation} The reaction wheels-based control follows the expression:

\begin{equation}
    \label{eq:RW based control}
    \mathbf{M}^{\mathcal{B}} = -\dot{\mathbf{H}} - \boldsymbol{\omega}^{\mathcal{B}}\times\mathbf{H},
\end{equation}

\noindent where $\mathbf{H} = J_{\mathrm{rw}}\mathbf{A}_{\mathrm{rw}}\boldsymbol{\Omega}$ is reaction wheels' assembly relative angular momentum, $\boldsymbol{\Omega} = [\Omega_1, \Omega_2, \Omega_3, \Omega_4]^{\top}$, $\dot{\mathbf{H}} = J_{\mathrm{rw}}\mathbf{A}_{\mathrm{rw}}\dot{\boldsymbol{\Omega}}_{\mathrm{ctrl}}$.

The torque allocation is done using conventional minimum norm pseudo-inverse method~\cite{MarkleyCrassidis2014}  and taking into account reaction wheel acceleration saturation:

\begin{subequations}\label{eq:rwa_allocation}
\begin{align}
\dot{\boldsymbol{\Omega}}_{\mathrm{ctrl}} &=
\mathbf{A}_{\mathrm{rw}}^{\top}
\left(\mathbf{A}_{\mathrm{rw}}\mathbf{A}_{\mathrm{rw}}^{\top}\right)^{-1}
(-\mathbf{u}_{\mathrm{ctrl}}) J_{\mathrm{rw}}^{-1}
\label{eq:rwa_allocation_nominal}
\\
\dot{\boldsymbol{\Omega}}_{\mathrm{ctrl}} &=
\begin{cases}
\dot{\boldsymbol{\Omega}}_{\mathrm{ctrl}},
& \max\!\left(|\dot{\Omega}_{\mathrm{ctrl},i}|\right) \le \dot{\Omega}_{\mathrm{max}} \\[6pt]
\dfrac{\dot{\Omega}_{\mathrm{max}}}
{\max\!\left(|\dot{\Omega}_{\mathrm{ctrl},i}|\right)}
\dot{\boldsymbol{\Omega}}_{\mathrm{ctrl}},
& \max\!\left(|\dot{\Omega}_{\mathrm{ctrl},i}|\right) > \dot{\Omega}_{\mathrm{max}}
\end{cases}
\label{eq:rwa_allocation_saturation}
\end{align}
\end{subequations}

\noindent where $\dot{\Omega}_{\mathrm{max}}$ denotes the maximum allowable reaction wheel acceleration according to the probe specifications provided in Section~\ref{sec: Lander description}. Attitude determination relies on a pair of sun sensors and an asteroid sensor. The sensors models are not considered in the study and will be investigated in further research.

\paragraph{Attitude Dynamics and Control Modeling}

\begin{figure}[h]
 \begin{center}
  \includegraphics[width=1\textwidth]{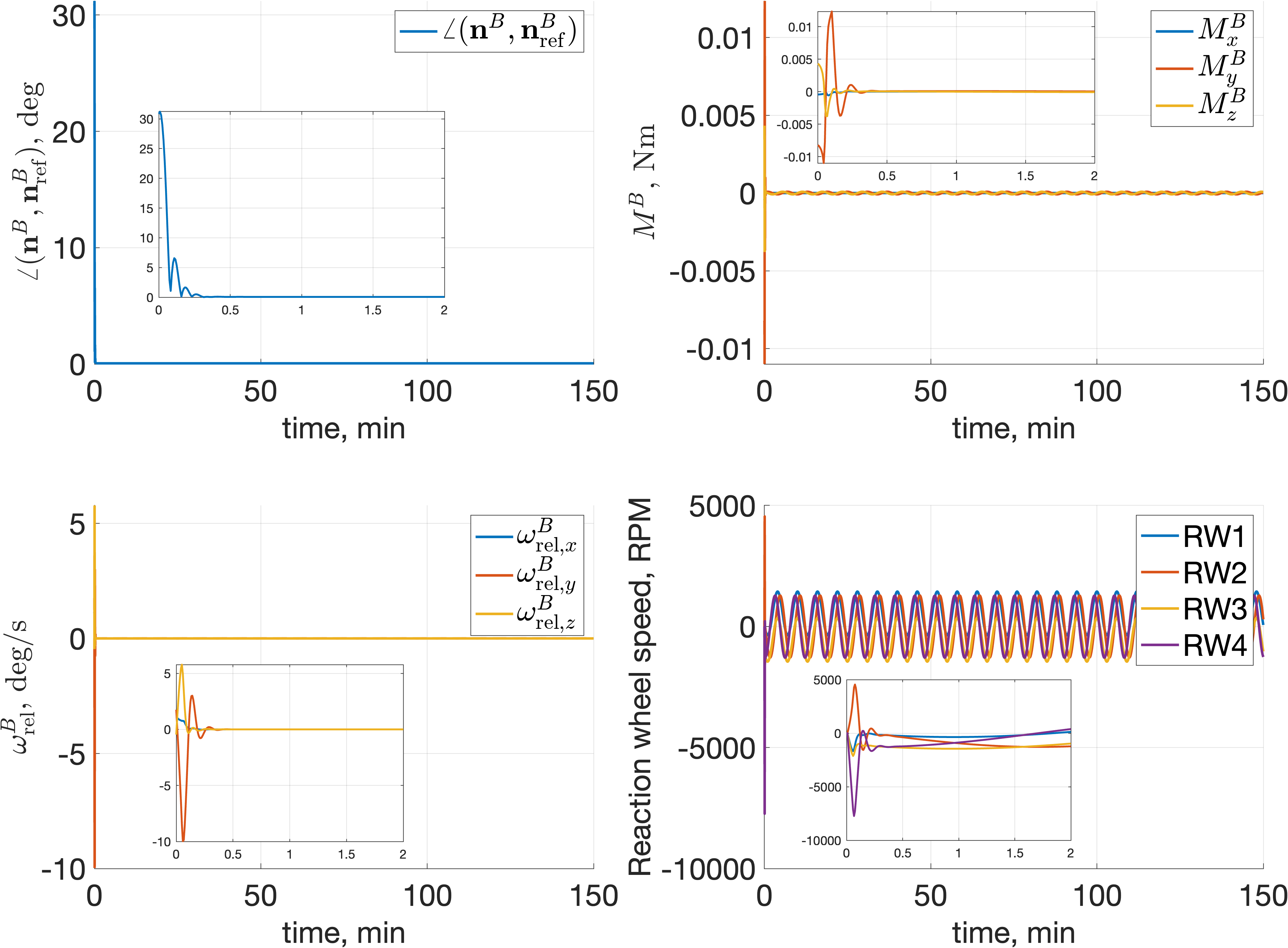}
  \caption{Sun-spin attitude control performance for the Mid-$\mu$ descent case with a target spin rate of 1~deg/s: Sun-pointing angular error, relative angular velocity error, control torque components, and reaction wheel speeds
}
  \label{fig:adcs sun spin}
   \end{center}
\end{figure}

\begin{figure}[h]
 \begin{center}
  \includegraphics[width=1\textwidth]{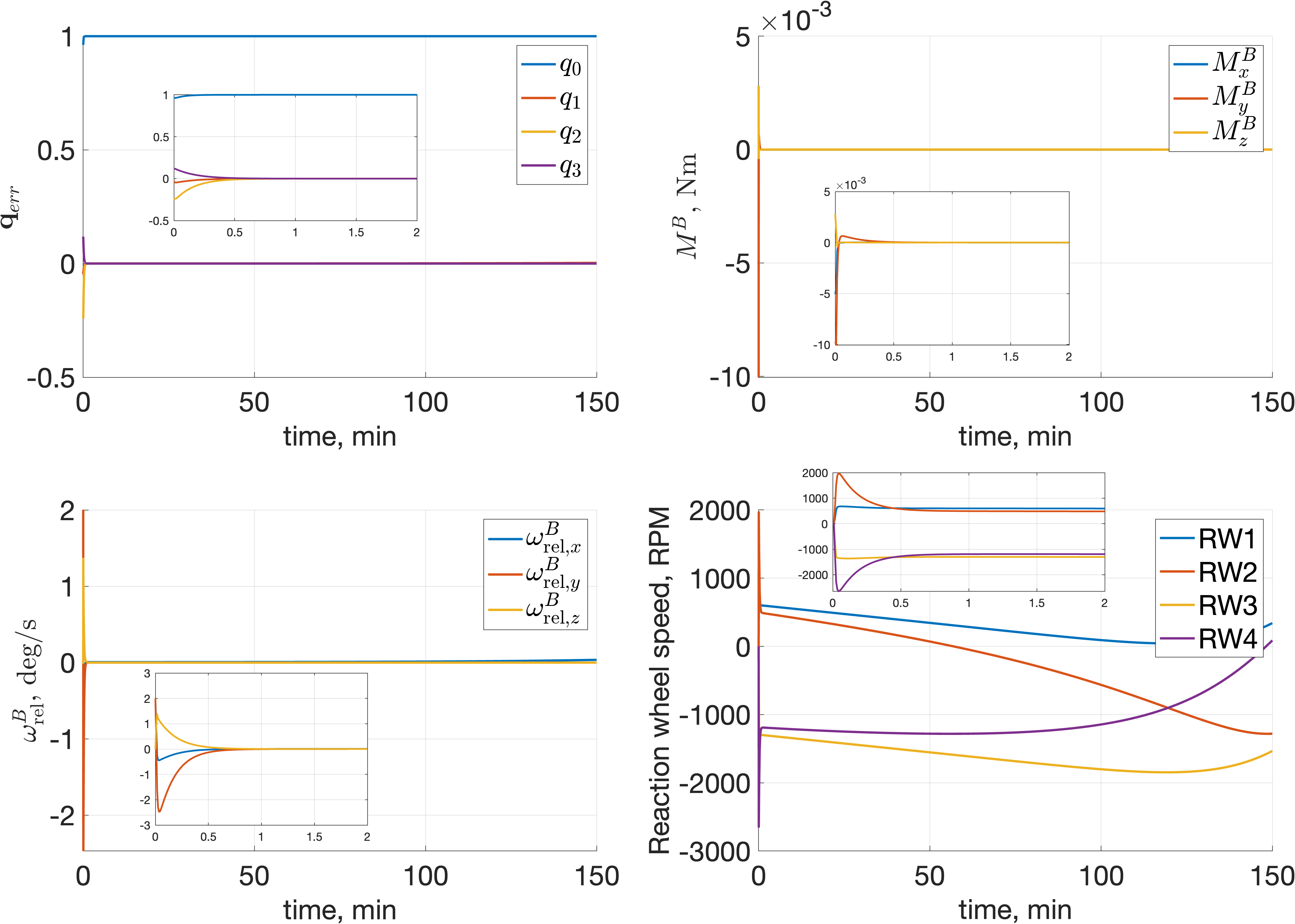}
  \caption{Asteroid-pointing attitude control performance for the Mid-$\mu$ descent case: attitude error unit quaternion, relative angular velocity, control torque components, and reaction wheel speeds
}
  \label{fig:adcs asteroid pointing}
   \end{center}
\end{figure}

The attitude dynamics and control modeling is performed together with the translational descent dynamics, since the control laws require the instantaneous Sun and asteroid direction vectors along the trajectory. The Mid-$\mu$ case is used as a representative scenario. Figures~\ref{fig:adcs sun spin} and~\ref{fig:adcs asteroid pointing} show the tracking performance of the Sun-spin and asteroid-pointing modes, respectively, together with the corresponding control torque vector and reaction wheel speed evolution.

In the Sun-spin case, the plotted tracking quantities are the angular deviation between the controlled body-fixed Sun-pointing direction and the reference Sun direction, and the relative angular velocity or angular velocity error with respect to the prescribed 1~deg/s spin rate around the Sun direction vector. In the asteroid-pointing case, the attitude tracking is characterized by the error quaternion computed from the attitude error matrix $\mathbf{C}_{\mathrm{err}}$, together with the relative angular velocity error. The control torque components $\mathbf{M}^\mathcal{B}$ and reaction wheel speeds are shown to illustrate the actuator response during the initial transient and the subsequent steady tracking phase. The results show that both controllers converge after the initial transient and maintain the required target states throughout the ballistic descent.

\end{document}